\documentclass{aa}  

\usepackage{graphicx}
\usepackage{txfonts}
\usepackage{hyperref}           
\usepackage[]{subfig}
\usepackage{longtable}
\usepackage{color}
\newcommand{\kms}   {~km~s$^{-1}$}
\newcommand{\ta}    {$T_{\mathrm a}^*$}
\newcommand{\tmb}   {$T_{\rm mb}$}

\begin{document} 

   \title{A spectroscopic survey of Orion KL between 41.5 and 50 GHz}

   \author{J. R. Rizzo\inst{1}
          \and
           B. Tercero\inst{2}
          \and
           J. Cernicharo\inst{2}
                     }

   \institute{Departamento de Astrof\'{\i}sica. Centro de Astrobiolog\'{\i}a (INTA-CSIC). 
              Ctra.~M-108, km.~4, 
              E-28850 Torrej\'on de Ardoz, Spain\\
              \email{ricardo.rizzo@cab.inta-csic.es}
         \and
             Grupo de Astrof\'isica Molecular. Instituto de Ciencia de 
             Materiales de Madrid (CSIC). Calle Sor Juana In\'es de la Cruz 3, 
             Cantoblanco, E-28049 Madrid, Spain}

   \date{Received October 21, 2016; accepted May 20, 2017}

\abstract
{The nearby massive star-forming region Orion KL is one of the richest 
molecular reservoirs known in our Galaxy. The region hosts newly formed 
protostars, and the strong interaction between their radiation and their 
outflows with the environment results in a series of complex chemical processes 
leading to a high diversity of interstellar tracers. The region is therefore 
one of the most frequently observed sources, and the site where many molecular 
species have been discovered for the first time.
}
{With the availability of powerful wideband backends, it is nowadays possible 
to complete spectral surveys in the entire mm-range to obtain a 
spectroscopically unbiased chemical picture of the region.
}
{In this paper we present a sensitive spectral survey of Orion KL, made with 
one of the 34~m antennas of the Madrid Deep Space Communications Complex in 
Robledo de Chavela, Spain. The spectral range surveyed is from 41.5 to 50\,GHz, 
with a frequency spacing of 180~kHz (equivalent to $\approx1.2$~km\,s$^{-1}$, 
depending on the exact frequency). The {\it rms} achieved ranges from 8 to 
12~mK.
}
{The spectrum is dominated by the $J=1\rightarrow0$ SiO maser lines and by 
radio recombination lines (RRLs), which were detected up to $\Delta n=11$. 
Above a $3\,\sigma$ level, we identified 66 RRLs and 161 molecular lines 
corresponding to 39 isotopologues from 20 molecules; a total of 18 lines remain 
unidentified, two of them above a $5\,\sigma$ level. Results of radiative 
modelling of the detected molecular lines (excluding masers) are presented.
}
{At this frequency range, this is the most sensitive survey and also the one 
with the widest band. Although some complex molecules like CH$_3$CH$_2$CN and 
CH$_2$CHCN arise from the hot core, most of the detected molecules originate 
from the low temperature components in Orion KL. 
}

\maketitle
%

\section{Introduction}
Orion KL has been widely recognized as the nearest high-mass star-forming 
region at a distance of 414\,$-$\,418\,pc \citep{men07, kim08}, and one of the 
richest molecular reservoirs known in the Galaxy. 

It hosts newly formed protostars, with strong interaction between their 
radiation and their outflows with the environment. Diverse chemical processes 
are therefore favoured in the region, which results in a high variety of 
molecules detected. Indeed, this is the site where many molecular species have 
been discovered for the first time \citep[e.g. methyl acetate,][]{ter13}. After 
several observational works \citep[e.g.][]{bla87,sch01,ter10,gon15}, at least 
four different molecular components have been identified in a relatively small 
volume. There are two compact sources, the \textit{hot core}, characterized by 
high temperatures ($T_{\rm K}$\,$\sim$100$-$300\,K) and the 
\textit{compact ridge}, with $T_{\rm K}$\,$\sim$100$-$150\,K. The third 
component is a quite extended \textit{plateau} (with an angular size of up to 
30\arcsec), characterized by a warm molecular outflow at 
$T_{\rm K}$\,$\sim$150\,K). The hot core, the compact ridge and the outflow are 
immersed within the cooler ($T_{\rm K}$\,$\sim$40$-$60\,K) \textit{extended 
ridge} or ambient cloud . These gas components could be resolved with single 
dish telescopes because they display different radial velocities and line 
widths. The Orion KL region is therefore an excellent testbed for the search of 
new molecules and also for the chemical characterization of those already known.

Many spectral line surveys have been carried out toward Orion KL covering a 
large range of frequencies (see Table\,1 of \citealt{gon15}). In recent years, 
we can mention the IRAM 30m millimeter survey \citep{ter10}, the Herschel/HIFI 
submillimeter and Far IR survey \citep{cro14}, the submillimeter survey carried 
out with the Odin satellite \citep{olo07,per07}, and the Effelsberg-100m 
telescope 1.3\,cm survey \citep{gon15}.

However, the spectral range around 7\,mm remains almost unexplored. At the best 
of our knowledge, the most complete survey up to date is that of 
\citet{god09a}, which covers a frequency range optimized for the 
$J=1\rightarrow0$ SiO maser emission lines (from 42.3 to 43.6 GHz). 

Besides the SiO lines, the $40-50$\,GHz spectral window is potentially rich in 
several complex organic molecules, and hosts several transitions also 
identified at higher frequencies (3, 2, and 1\,mm), but at lower energy levels. 
At these frequencies the presence of a wealth of radio recombination lines 
(RRLs) is also remarkable . 

In this paper we present the results of a spectral survey in the Orion KL 
region, in the frequency range from 41.5 to 50~GHz (6 to 7.2\,mm in 
wavelength). Sect.~2 describes the technical aspects of the survey, while 
Sect.~3 summarizes the results. A radiative model of the molecular emission is 
presented in Sect.~4, and the conclusions are enumerated in Sect.~5.

%

\section{The spectral survey}
We used the DSS-54 antenna, one of the 34m dishes available at the NASA's 
Madrid Deep Space Communications Complex (MDSCC), to perform a sensitive 
spectral survey of Orion KL. The observations were done in different runs from 
December 2013 to February 2014.

The survey was performed using a cooled high electron mobility transistor 
(HEMT) receiver \citep{riz13} and the two circular polarizations have been 
recorded. The receiver temperature remained around 40\,K. The resulting system 
temperature varied from 90 to 190~K (in antenna temperature units), depending 
on the frequency, elevation and weather conditions. The backend was a new 
wideband backend \citep{riz12}, which provides 1.5 GHz of instantaneous 
bandwidth and a resolution of 180~kHz (equivalent to $\approx 1.2$~km~s$^{-1}$ 
at the observed frequencies), for each circular polarization.

The survey was conducted in position switching mode in six sub-bands, with a 
superposition of 100 MHz between two consecutive sub-bands, in order to check 
consistency and eliminate possible image sideband effects. Total integration 
time was 1490 minutes (on source). For each sub-band the integration time 
varied from 97 to 422 minutes, in order to reach to a uniform $rms$ (1-sigma) 
level between 4 and 6~mK, on an antenna temperature scale. 

Observations were carried out in winter nights, under good weather conditions. 
The atmospheric opacity has been measured during each observing session by means 
of tipping curves, and resulted to be between 0.07 and 0.11. During the 
observations, individual spectra have been corrected by atmospheric opacity and 
antenna gain due to elevation, obtaining the antenna corrected scale (\ta).

Data were processed using CLASS, a part of the GILDAS software\footnote{GILDAS 
is a radio astronomy software developed by IRAM. See 
{\tt http://www.iram.fr/IRAMFR/GILDAS}/.}. Spectra from different runs were 
averaged and baseline subtracted. The final spectrum was converted to main beam 
brightness temperature (\tmb) according to the usual expression 

\begin{equation}
T_{\rm mb} = T_{\rm a}^* / \eta_{\rm mb},
\end{equation}

\noindent 
where $\eta_{\rm mb}$ is the main beam efficiency \citep{wil09}. The 
most relevant antenna parameters for different scale conversions, including 
density flux $S$, are summarized in Table \ref{antenna}. Unless otherwise 
specified, we use the \tmb\ scale throughout the paper.

%

\section{Results}
\subsection{Overall emission}
\begin{figure*}
      \centering
      \includegraphics[width=\textwidth]{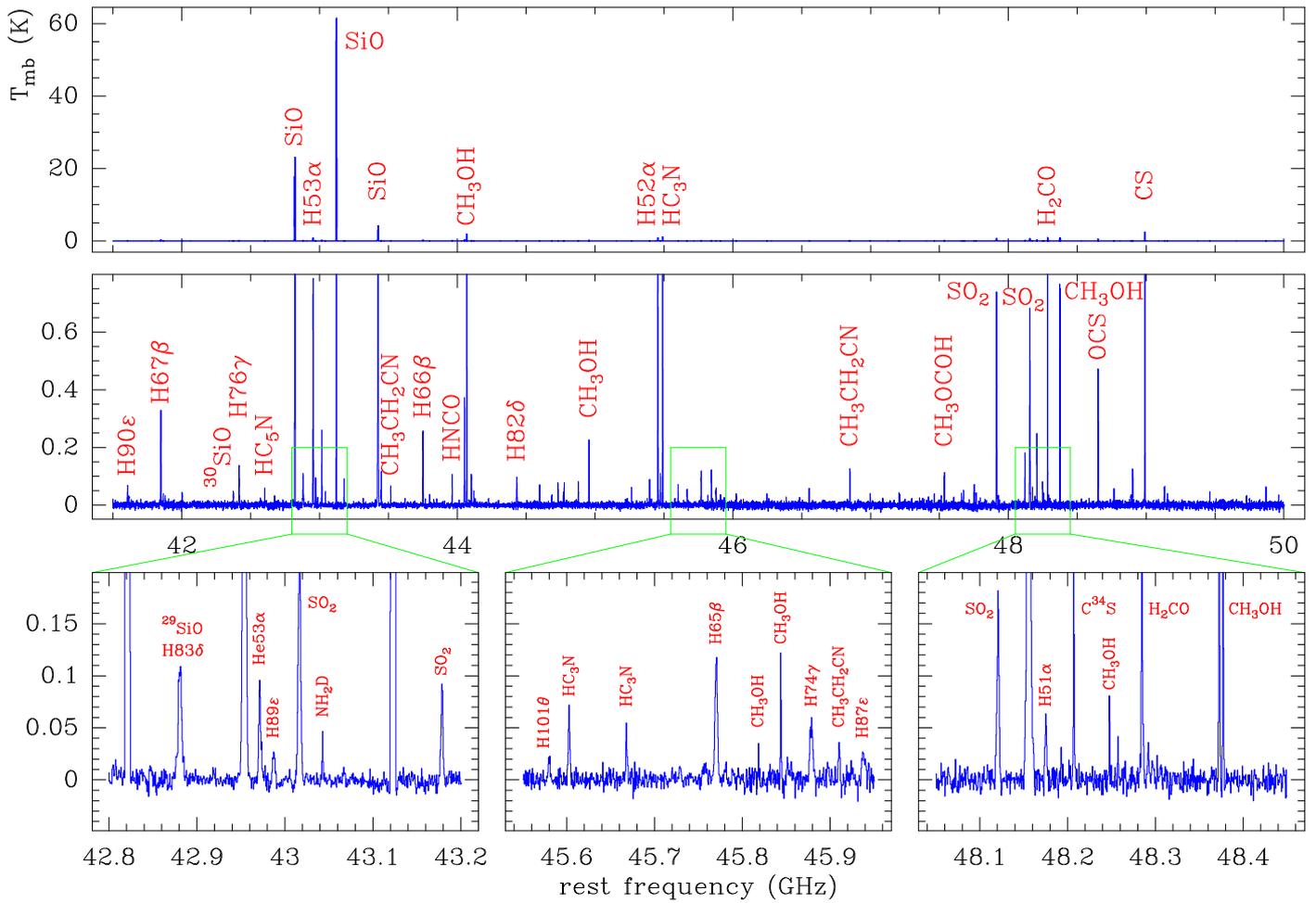}
      \caption{Spectrum of Orion KL. Top panel: full range. Middle panel: 
      zoom in intensity to see low intensity lines; green boxes are expanded 
      in the three bottom panels, to show spectroscopic details. Some of the 
      most intense lines are labelled.
              }
         \label{allrange}
   \end{figure*}

Figure \ref{allrange} shows the resulting spectrum in the whole observed band. 
The upper panel displays the full range both in frequency and in intensity. We 
see that the spectrum is dominated by the emission of SiO masers and RRLs. Some 
well known molecules (CS, H$_2$CO, HC$_3$N, CH$_3$OH) are also intense and 
usually associated with both, cold and hot gas.

The middle panel depicts a zoom in intensity, where it is possible to 
distinguish a number of other, more complex molecules, particularly SO$_2$, 
CH$_3$CH$_2$CN, HNCO, CH$_3$OCOH, and OCS. As shown in the model (Sect.~4), 
these molecules are associated with the hottest parts of the region.

The three lower panels are small spectral windows (indicated in green in the 
middle panel) which display the richness of spectral lines and different line 
widths. We identified a total of 66 RRLs and 161 molecular lines above a 
$3\,\sigma$ level, of which 18 lines remain unidentified (U lines hereafter). 
The relative number of molecular lines with respect to RRLs is therefore 
$161/66\approx2.4$. This is $\approx4$ times higher than the ratio obtained at 
1.3~cm \citep[for example][who detected 164 RRLs and 97 molecular 
lines]{gon15}, but notably lower than that found at millimeter wavelengths, 
where \citet{ter10} identified more than 14400 molecular lines but only a few 
dozens of RRLs. 

Therefore, the 7~mm band is especially useful in carrying out spectroscopic 
studies where the emission of RRLs and molecular lines are both of interest. 
Within some GHz of bandwidth, this band has a well balanced number of RRLs and 
molecular lines. At cm wavelengths, the number of molecular lines is quite low, 
while at mm wavelengths the high density of molecular lines and the large 
separation in frequency of the RRLs makes it difficult to simultaneously study 
the atomic ionised and the molecular components.

The procedure employed for line identification (such as smoothing, catalogues, 
and cross-check with other surveys) is outlined in Sect.~4. 

\subsection{Radio recombination lines}
The detected 66 RRLs include 50, 12, and 4 lines corresponding to hydrogen, 
helium, and carbon, respectively. Hydrogen has been detected at a maximum 
$\Delta n$ (the difference between principal quantum numbers of the upper and 
lower levels) of 11. All the detected lines can be well characterized by 
Gaussian fittings, whose parameters are indicated in Table~\ref{rrl}. For each 
detection, the table contains the line name (ordered by increasing $\Delta n$), 
frequency, velocity-integrated line intensity, peak velocity, full width at 
half maximum, and peak temperature. One-sigma errors are displayed within 
parenthesis. Frequencies have been obtained from the MADCUBA\_IJ package. 
\footnote{{\tt http://cab.inta-csic.es/madcuba/}.}

Possible blending with molecular lines or other RRLs are indicated as notes in 
the last column of Table~\ref{rrl}. In some cases (H64$\beta$, H83$\delta$, and 
H107$\kappa$) we could not separate the contribution of the blending lines, and 
the corresponding fitting has not been considered in subsequent analysis. The 
line H81$\delta$ is clearly detected, but the spectrum is suffering from a 
series of unwanted features in specific channels (``spikes'') at close 
frequencies; therefore, the fitting seriously underestimates the line intensity 
(probably up to a factor of 3).

\begin{figure}
      \centering
      \includegraphics[width=\columnwidth]{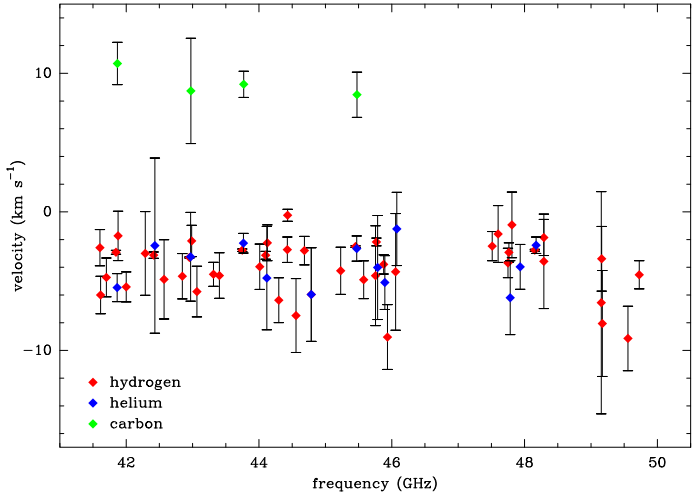}
      \caption{Distribution of the velocities of the RRLs as a function of 
      frequency. Atoms (hydrogen, helium, and carbon) are labelled in red, 
      blue, and green, respectively.
              }
         \label{velocities}
   \end{figure}

For each atom, Fig.~\ref{velocities} depicts the distribution of velocities as 
a function of frequencies. The unweighted mean velocities of hydrogen and 
helium agree within uncertainties, with values of $-4.0\pm1.2$~\kms\ for 
hydrogen and $-3.6\pm1.8$~\kms\ for helium, while for the carbon lines we 
obtain $+9.3\pm1.9$~\kms. To compute these values we included all the lines 
detected not affected by blending, and the error was estimated as the maximal 
individual error divided by the square root of the number of lines. The 
corresponding mean velocities weighted by the inverse square of the  individual 
errors are $-2.9\pm0.1$, $-3.0\pm0.4$, and $+9.4\pm0.7$~\kms\ for hydrogen, 
helium and carbon, respectively. As expected, these values are dominated by the 
$\alpha$ lines which have the smallest uncertainties. The computed velocities 
are compatible with hydrogen and helium arising from the {\sc Hii} region M42, 
while the carbon recombination lines originate from the photon-dominated region 
(PDR) at the interface between M42 and the associated molecular cloud 
\citep[the Orion Bar, see][]{cua15,gon15}. 

Based on their own data and other works, \citet{god09a} reported that RRL 
velocities increase with frequency, and explain this effect by the presence of 
a density gradient within the {\sc Hii} region. We decided to explore such 
hypothesis by a careful reading of the available bibliography. First, 
\citet{god09a} report ``{\it velocities ranging from $-$5 to $-$10\kms}'' based 
on three RRLs (H53$\alpha$, He53$\alpha$, and H76$\gamma$), but they do not 
provide individual fitting; we measure a mean velocity of the H53$\alpha$ line 
of $-3.29\pm0.04$\kms. Second, the other observations used for the hypothesis 
are those at 71$-$122~GHz from \citet{tur91} (18 RRLs, including four $\alpha$ 
lines from hydrogen) and at 215$-$247~GHz from \citet{sut85} (2 RRLs, only one 
$\alpha$ line from hydrogen). \citet{tur91} reported a mean velocity of 
$-$3.4\kms, but unfortunately did not provide individual fitting of the lines. 
\citet{sut85} reported $+$4\kms\ for the H30$\alpha$ line, but this is clearly 
contaminated by a rather strong line of $^{33}$SO$_{2}$ at 
$\approx$231.9003~GHz \citep{esp13a}.

\begin{figure}
      \centering
      \includegraphics[width=0.9\columnwidth]{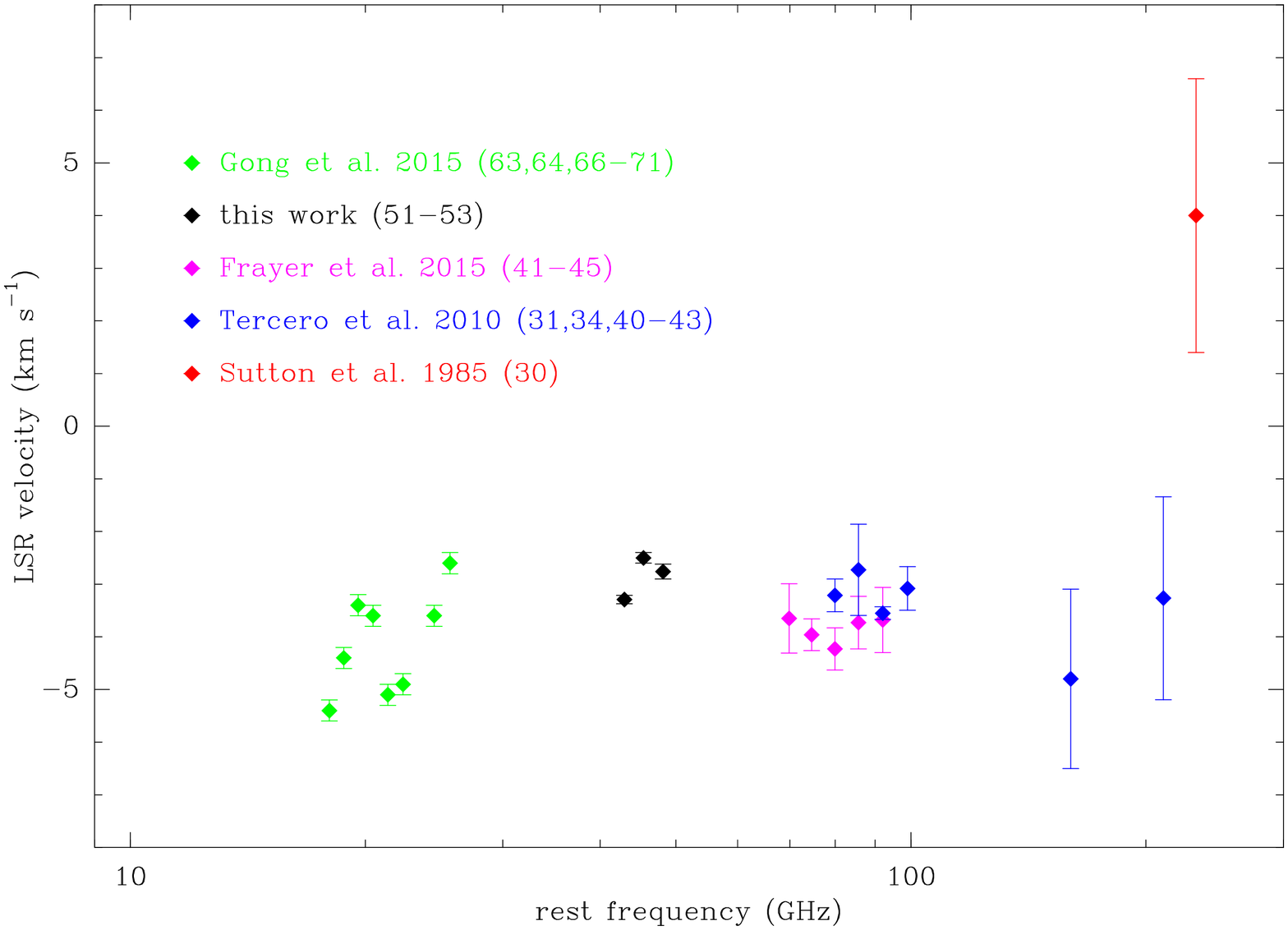}
      \caption{LSR velocities of $\alpha$ RRLs from hydrogen in Orion KL, as a 
               function of the frequency (on a logarithmic scale). Surveys used 
               are indicated with different colours. Together with the 
               references, the principal quantum numbers are indicated within 
               parenthesis. Error bars correspond to 2-sigma values.
              }
         \label{Halpha-veloc}
   \end{figure}

After the work of \citet{god09a}, several new surveys have been performed at 
different wavelengths. We therefore searched for the available information 
about $\alpha$ RRLs from hydrogen. Besides our own survey, we selected those 
works providing individual fitting \citep{sut85,gon15}, or with the possibility 
to gather frequency calibrated data, as the IRAM 30m data provided by 
\citet{ter10} and the GBT 4~mm survey of \citet{fra15}, available 
online\footnote{{\tt 
http://vizier.cfa.harvard.edu/viz-bin/VizieR?-\\source=J/AJ/149/162}}. 
The results are summarized in Table~\ref{Halpha-tab}, and depicted in 
Fig.~\ref{Halpha-veloc}. Data include Gaussian fitting of $\alpha$ hydrogen RRL 
velocities from H30$\alpha$ to H71$\alpha$, covering a wide range of 
frequencies from 17 to 232~GHz. We do not see in the figure or the table any 
significant variation of the velocities at any frequency, except the 
H30$\alpha$ line, which is affected by blending, as already commented above. 
Therefore, we have not found any evidence for the claimed density gradient in 
the {\sc Hii} region.

Line widths are also quite uniform for each atom. We computed mean values of 
$21.9\pm3.4$, $16.4\pm6.2$, and $5.4\pm3.4$~\kms\ for hydrogen, helium and 
carbon, respectively, which are in agreement with previous results 
\citep{gon15}. 

\begin{figure}
      \centering
      \includegraphics[width=\columnwidth]{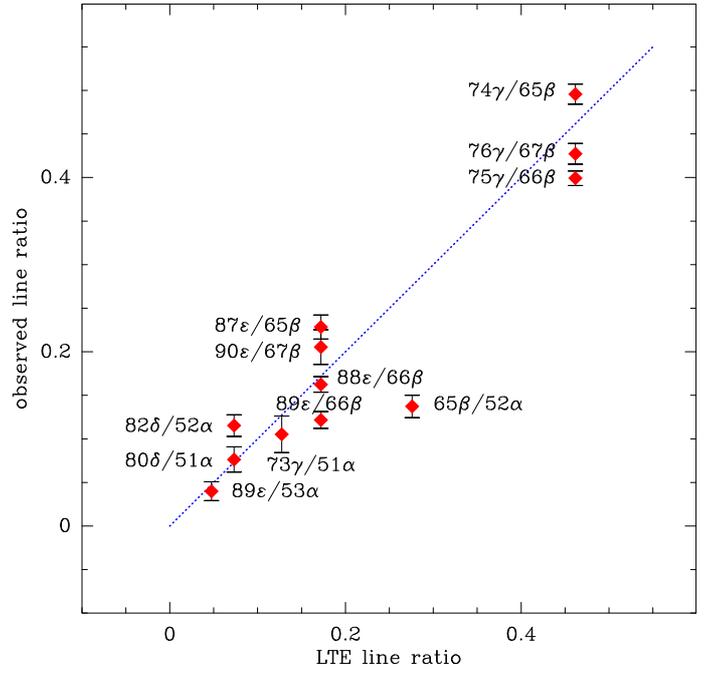}
      \caption{Distribution of the measured line intensity ratios between pairs 
      of hydrogen RRLs at nearby frequencies as a function of the predicted 
      ratios under LTE conditions. The ratios are labelled and the bars 
      represent 1.5-sigma errors. The dotted blue line corresponds to equal 
      values between observed and predicted ratios under LTE conditions.
              }
         \label{LTE}
   \end{figure}

In order to test the physical conditions of the {\sc Hii} region, we computed 
the ratio of peak temperatures for two hydrogen RRLs which lie at close 
frequencies. These ratios were then compared to the expected values under LTE 
conditions, following the procedure outlined by \citet{bro72}. We excluded the 
ratios involving the H81$\delta$ and H83$\delta$ lines (see notes in 
Table~\ref{rrl}). The result is depicted in the Fig.~\ref{LTE}, where the 
measured ratios are plotted as a function of the LTE ratio. We see in this 
figure that the excitation of RRLs at these frequencies are close to LTE, 
within the uncertainties.

We also measured the helium abundance by computing the velocity-integrated line 
ratio between the same RRL from helium and hydrogen. We used all the available 
$\alpha, \beta,$ and $\gamma$ lines, excluding H64$\beta$ due to blending. 
These ratios are a measure of the helium abundance under LTE conditions, 
assuming that all the helium is singly ionized and both Str\"omgren spheres are 
identical. The resulting value is $8.3\pm1.2$~\%, which agrees with previous 
measurements in this source \citep{gon15}, and is close to Solar System values 
\citep{wil94}.

\subsection{SiO emission}

\begin{figure}
      \centering
      \includegraphics[width=0.9\columnwidth]{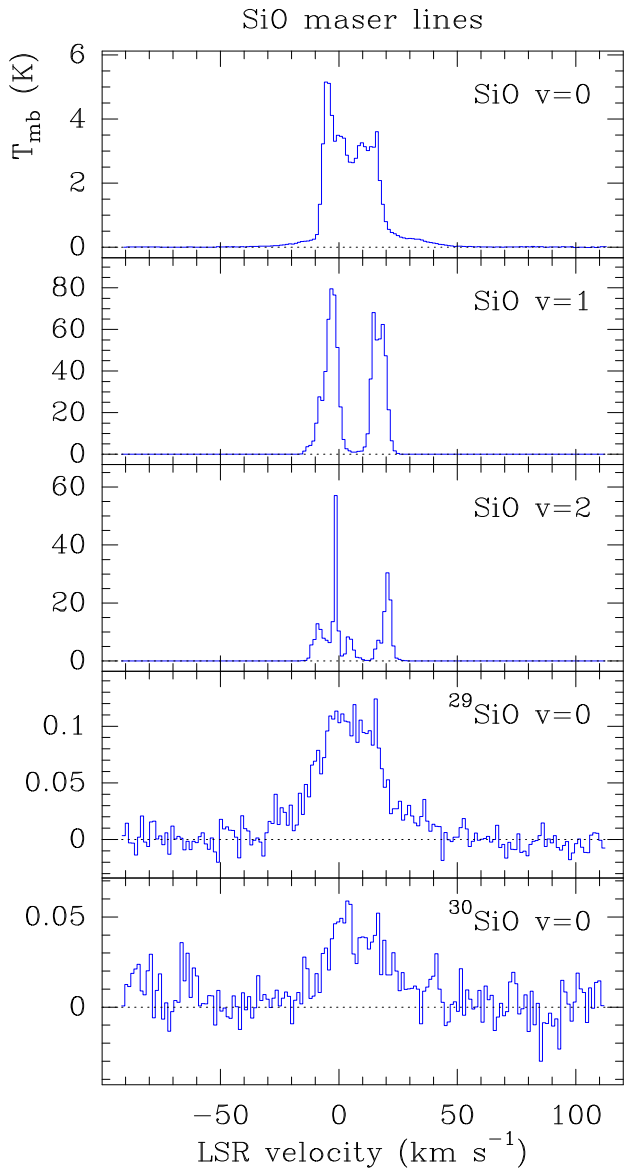}
      \caption{The SiO $J=1\rightarrow0$ lines detected in Orion KL. 
              Isotopologues and vibrational numbers are indicated in each panel.
              }
         \label{SiO}
   \end{figure}

We detected a total of 5 lines from SiO, which are shown in Fig.~\ref{SiO}. The 
lines correspond to the vibrational numbers $v=0$, 1, and 2 of the main 
isotopologue, and to $v=0$ of $^{29}$SiO and $^{30}$SiO. Overall, the line 
shapes agree with the single-dish results presented by \citet{god09a}.

The most intense lines are masers corresponding to $v=1$ and 2 of the main 
isotopologue. These lines are double-peaked and cover the velocity range from 
$-$16 to +27\kms. \citet{god09b} have mapped these lines using the Very Large 
Array (VLA) and found that the emission arises from the inner part of the 
circumstellar structure associated with source I, with a maximum size of 
100~AU.

The bright and very narrow feature in the $v=2$ line at $\approx -1.4$\kms\ is 
reported here for the first time. This component is not present in previous 
observations of \citet{god09a,god09b}, obtained with even better spectral 
resolution. The lack of a similar feature in the $v=1$ line may indicate that 
the emission of both maser lines arises from slightly different regions, as 
\citet{god09b} suggested.

The bulk of the $v=0$ emission from $^{29}$SiO and $^{30}$SiO lines appears as 
a single component matching approximately the same velocity range as the $v=1$ 
and 2 lines of $^{28}$SiO. The high resolution maps \citep{god09b} show that 
the emission of the four lines is confined to a distance of $\approx100$~AU to 
a young stellar object known as source~I. \citet{god09b} also pointed out 
significant variations (at least two orders of magnitude) in the brightness 
temperature. The above facts and further modelling of radiative pumping allow 
the authors to conclude that the $^{29}$SiO and $^{30}$SiO $v=0$ line emission 
is non thermal.

The line shape of the $v=0$ line from $^{28}$SiO is similar to those of 
$^{29}$SiO and $^{30}$SiO, although it has a broader spatial distribution 
\citep{god09b}, probably due to an outflow or a disc with a size of 
$\approx 600$~AU in size, produced in the last $10^3$~yr \citep{ter11,nie12}.

The emission of the ground vibrational state of $^{28}$SiO has been observed at 
several rotational transitions up to $8\rightarrow 7$ 
\citep[e.g.][]{beu05,ter11,nie12}. All these lines except $1\rightarrow 0$ have 
Gaussian-like shapes and do not show hints of maser emission 
\citep[in fact,][satisfactorily modelled their emission under thermal 
regimes]{ter11}. The shape of the $v=0$ $J=1\rightarrow0$ spectrum, as shown 
here, is rather different from the other $v=0$ lines of $^{28}$SiO and the 
thermal modelling of \citet{ter11} is not able to reproduce the observed shape 
and intensity. Therefore it seems probable that the line is caused at least 
partially by maser emission, which has also been proposed by \citet{cha95}.

In addition to the main velocity component, the $v=0$ lines of the three 
isopotologues have large wings from $-$40 to +60\kms. The wings have been 
mapped in other rotational lines using interferometers: 
\citep[$J=1\rightarrow 0$:][VLA]{god09b}, 
\citep[$J=2\rightarrow 1$:][CARMA]{pla09}, and \citep[$J=5\rightarrow 4$, 
$6\rightarrow 5$:][ALMA]{zap12,nie12}. All those high resolution observations 
show that this wide component arises from a bipolar outflow driven by source I 
\citep{god09b}.

%

\section{Radiative model of the molecular emission}
\subsection{Overall description}

Figure~\ref{fig_model} displays the observed spectrum with more detail, after 
applying Hanning-smoothing of four contiguous channels, i.e., after degrading 
the spectral resolution to $\sim$0.7\,MHz. The intensity scale was in some 
cases zoomed to improve the visibility of weak lines (grey and yellow boxes of 
Fig.~\ref{fig_model}). Superimposed on the observed spectrum is also the 
result of the line identification and further modelling. The SiO species and 
RRLs, although identified, are not part of the model.

The complete list of the identified species is presented in 
Table~\ref{table_linelist}. The frequencies have been obtained from the 
CDMS\footnote{{\tt http://www.astro.uni-koeln.de/cgi-bin/cdmssearch}.} 
\citep{mul01,mul05}, the 
JPL\footnote{{\tt http://spec.jpl.nasa.gov/ftp/pub/catalog/catform.html}.} 
\citep{pic98}, and private catalogues (J. Cernicharo, private communication). 
In all cases, we selected the frequency having the highest precision or that 
which provides the most complete information about hyperfine components.
Most of the lines have been successfully identified, with only two U lines at a 
5$-$6$\sigma$ level, at 46387.9 and 49765.7\,GHz. The first U line has a width 
of $\approx$ 10\kms, while the second one has $\approx$ 20\kms. If confirmed, 
the 46387.9\,GHz line may arise from the hot core and the other one from the 
plateau. Due to the low signal-to-noise ratio, it is hard to provide a robust 
characterization of the line shapes. All lines identified in this survey 
correspond to abundant molecules previously detected in this source 
\citep[e.g.][]{bla87,sut95,sch01,god09a,ter10,ter11}.

For many of the identified species, we previously derived physical parameters 
and column densities based on data from the IRAM line survey at 3, 2 and 
1.3\,mm \citep{ter10}, which have been already published in a series of papers 
(see Table B3 of \citealt{cer16} for references). Due to the wide frequency 
range of that survey and the detailed analysis of the different components, we 
expect that the previous results for the modelled species are well constrained. 
Therefore, as a first approach we applied the models already performed to fit 
the lines of the IRAM 30m line survey to model the molecular emission at 7\,mm.

We have built new radiative models for those molecules lacking previous 
modelling at this or other frequency bands after considering the differences 
attained from the use of different telescopes and frequencies. We fitted 
simultaneously both the cold and hot components in all the species. To do so, 
we used the MADEX code \citep{cer12}, which solves simultaneously the radiative 
transfer and the statistical equilibrium equations using local thermodynamic 
equilibrium (LTE); when the collisional rates were available, we proceeded with 
the large velocity gradient approach (LVG). A total of 5020 species from 1107 
families\footnote{We call family to a group of molecules containing all the 
isotopologues of a given species and all the vibrationally excited molecules, 
like for example SiO, $^{29}$SiO, $^{30}$SiO, and SiO $v=1, 2, ...$ }. can be 
modelled by this code. 

We introduced the physical parameters of the different components found in the 
source (kinetic temperature $T$$_{\rm K}$, H$_2$ volume density $n$(H$_2$), 
source diameter $d$$_{\rm sou}$, LSR velocity $v_{\rm LSR}$, and line width 
$\Delta$$v$) according to either the typical values found in the literature and 
our previous Gaussian analysis of the lines. Furthermore, we also have taken 
into account the telescope dilution and the relative offsets between each cloud 
component with respect to the pointing position. We have chosen the column 
density of each cloud component as the free parameter. The nature of this 
method is iterative, which allowed us to introduce corrections to the original 
values of $T_{\rm K}$ and $n$(H$_2$) in LVG calculations to improve the 
fitting. The discrepancies between the original and corrected values were below 
20\% in all cases.

At some frequencies, baseline fluctuations may produce differences in line 
intensity up to $\Delta$$T_{\rm mb}$\,$\sim$\,$\pm$0.02\,K. Therefore, the 
uncertainties may grow up to 20\% for those lines having an observed intensity 
below $T_{\rm mb}$\,=\,0.05\,K 

We estimated the uncertainties of the resulting column densities supported by 
the analysis already done in the IRAM 30m survey \citep{ter10}. There is not a 
remarkable new source of uncertainties in the new data presented in this work, 
especially for the molecules displaying a high signal-to-noise ratio in their 
lines. According to this analysis, the newly modelled molecules have 
uncertainties in the column density below 30\%.

The physical parameters and the column densities obtained by the models are 
summarized in Table~\ref{table_model}, where the contribution of every molecule 
to the different components is presented. Detailed results for each species are 
discussed in the next two subsections.

\subsection{Agreement with models for the IRAM 30m survey}

Most of the molecules with previous models have the bulk of their emission 
arising from the hot core. As described in the previous section, we used the 
physical parameters of those models as initial guesses. 

We satisfactorily reproduced the emission of some of these species (HCS$^+$, 
CCS, NH$_2$CHO, CH$_3$CH$_2$CN $\varv_{13}$/$\varv_{21}$, $^{34}$SO$_2$, 
$^{33}$SO$_2$, HC$_3$N $\varv_6$=1, and CH$_2$CHCN) without having to modify 
any parameters in the models \citep{ter10,mot12,dal13,esp13a,esp13b,lop14}. A 
look at the Table~\ref{table_model} shows that these molecules have minor or 
absent contributions from the extended ridge and are therefore well described 
by the models at 1, 2, and 3~mm. In the case of NH$_2$CHO, the upper limit to 
the column density in this survey is well constrained by the model of 
\citet{mot12}.

Some other species (CH$_3$CH$_2$CN, SO$_2$, $^{13}$CH$_3$OH, HC$_3$N, HC$_5$N, 
OCS, CS, C$_3$S, and HNCO) are more ubiquitous and have significant 
contribution from both the hot and the cold components. We have adapted the 
models of \citet{dal13} for CH$_3$CH$_2$CN, \citet{esp13a} for SO$_2$, 
\citet{kol14} for $^{13}$CH$_3$OH, \citet{esp13b} for HC$_3$N and HC$_5$N, 
\citet{ter10} for OCS, CS, and C$_3$S, and \citet{mar09} for HNCO in order to 
properly fit the lines of these species. In all cases, it was necessary to 
modify the column densities of the cold components with respect to the previous 
models. For SO$_2$, CH$_3$CH$_2$CN and $^{13}$CH$_3$OH we had to reduce the 
column densities of the extended ridge by a factor of two. 

The case of $^{13}$CS is the most outstanding, whose column dentisty has been 
reduced by up to a factor of ten in the cold component. This drastic change 
implies an intriguingly high $N$(CS)/$N(^{13}$CS) ratio (from 100 to 250), 
which should be studied through future and more detailed observations and 
models.

For C$_3$S we introduced for the first time an extended ridge component, which 
is evident in the $J$=8$\rightarrow$7 line (with upper level energy around 
10~K).

In the case of C$^{33}$S we detected only a narrow line, compatible with the 
extended ridge, which is by far the major contributor to the line profile. The 
warm components remain at an intensity level below 0.05~K, corresponding to the 
upper limits quoted in Table~\ref{table_model}. 

In short, we noted that the 7~mm lines are more sensitive to small variations 
in the column densities of the cold component than lines at shorter 
wavelengths. This survey is, therefore, more adequate to constrain the 
contribution of the extended ridge gas.

Although all species discussed here have been previously detected in this 
source, some of them are only tentatively identified near the detection limit 
of our data: NH$_2$CHO, O$^{13}$CS, CCS, $^{33}$SO$_2$, CH$_3$CH$_2$CN 
$\varv_{13}$/$\varv_{21}$, and HC$_3$N $\varv_7$=2 (see Fig.\,\ref{fig_model}). 
Therefore, the derived column densities for these species have to be considered 
as upper limits.

\subsection{New models}
   
\noindent {\bf CH$_3$OH (methanol).} A total of 10 lines of the main 
isotopologue of methanol are detected, equally distributed in A- and E- 
substates. Six of the lines are in the ground vibrational state, and four of 
them correspond to the first vibrationally excited state. The only maser is the 
$7_0\rightarrow6_1 A^+$ transition, which is a well known class I maser in 
several Galactic sources, including Orion KL \citep[e.g.][]{has90}. Orion~KL 
was, in fact, the first source where a methanol maser line was detected 
\citep{bar71}. The $7_0\rightarrow6_1 A^+$ line shape \citep{has90} is 
remarkably similar to the $8_0\rightarrow7_1 A^+$ line \citep{ohi86}, having a 
thermal component of $\approx4$\kms\ and a very narrow maser component with a 
width of only 0.22\kms\ or less. Unfortunately, the spectral resolution of this 
survey prevents us to separate both components. Models of methanol in the 
ground state and in its first vibrationally excited state have been performed 
considering the same components as those used for $^{13}$CH$_3$OH. We note that 
the abundance ratio [CH$_3$OH]/[$^{13}$CH$_3$OH] is between 5-15 for the 
different components. These values are far from the isotopic $^{12}$C/$^{13}$C 
ratio of 45 derived by \citet{ter10} in the source. Therefore, the lines from 
the main isotopologue of this abundant organic species are optically thick also 
at these frequencies. The derived column density for the hottest component 
(300\,K) of A-CH$_3$OH $\varv_t$=1 is a factor 10 higher than that of the 
E-substate. Only one line of A-CH$_3$OH $\varv_t$=1 with $E_{\rm u}$\,=\,426\,K 
appears in this survey above the detection limit whereas the energies of the 
lines of the E-species are around $\sim$300\,K (see 
Table~\ref{table_linelist}). Due to this difference in energy we obtain a 
different value for the column densities in the A- and E-substates. The low 
number of lines prevents a detailed analysis and we cannot accurately constrain 
the column densities.

\noindent {\bf H$_2$CO (formaldehyde).} $ortho$ and $para$ formaldehyde have 
been modelled for all the lines between 7\,mm and 1\,mm. The $ortho$-to-$para$ 
ratio is $\sim$3, as expected at high gas temperature. Only one line of 
o-H$_2$$^{13}$CO with low intensity ($S_{\rm ij}$$<$0.5) is expected in the 
present frequency range, at 45.9\,GHz. Because this line is only tentatively 
detected, the o-H$_2$$^{13}$CO column density has to be considered as an upper 
limit.

\noindent {\bf CH$_3$OCOH (methyl formate).} Ground state and vibrationally 
excited species ($\varv_{t}$=0,1) have been identified in the present 
data. For the model, we used the same components provided by \citet{mar10}. 
Lines corresponding to $b$-type transitions arising in the IRAM 30m data and 
the lines of the present survey have been fitted simultaneously. It is worth 
noting that the difference in the $a$ and $b$ dipole moments 
($\mu_{\rm a}$\,=\,1.63 and $\mu_{\rm b}$\,=\,0.68 from \citealt{cur59}) 
affects the opacity of the lines: many of the $a$-type transition-lines arising 
in the IRAM 30m survey are optically thick whereas those corresponding to 
$b$-type transitions are optically thin. By our procedure, we obtain a total 
column density of CH$_3$OCOH $\varv_t$=0 three times larger than the value 
provided by \citet{mar10} and \citet{car09}, who assumed LTE conditions and 
optically thin emission. The first vibrationally excited state has been 
detected by means of several faint lines, most of them near the detection 
limit. We derived the column density by reducing the column density values 
obtained for the ground state to fit these lines in the present survey.

\noindent {\bf CH$_3$OCH$_3$ (dimethyl ether).} We modelled this molecule by 
assuming that it is spatially correlated with methyl formate 
\citep{bro13,ter15}. We initially fitted the four components described in 
Sect.~1. Although two of the dimethyl ether lines have upper level energies as 
low as 2.3\,K and 9.1\,K, we had to remove the extended ridge to better 
reproduce the bulk of the dimethyl ether emission. The dominant component is by 
far the compact ridge, in agreement with high resolution observations 
\citep{bro13}.

\noindent {\bf NH$_2$D (deuterated ammonia).} We used the four typical 
components of Orion~KL (Sect.~1) to fit the NH$_2$D lines. We used the LTE 
approximation because only two lines are detected in our data. The intensity of 
the line at 49963 MHz results significantly overestimated, which suggests that 
the excitation of this species may be far from equilibrium.

\noindent {\bf CH$_2$OCH$_2$, CH$_3$COCH$_3$, CH$_3$NH$_2$ (ethylene oxide, 
acetone, methylamine).} Most of the lines associated to these molecules 
(Table~\ref{table_linelist}) are just tentatively detected (i.e. below the 
3$\sigma$ level) in this survey, and the derived column densities must be seen 
only as upper limits (Table~\ref{table_model}). We modelled these species by 
fitting the four components to the lines depicted in Table~\ref{table_linelist} 
and also to those clearly detected at 3, 2, and 1\,mm \citep{ter10}.

%

\section{Conclusions}
This is the most complete spectral survey of Orion KL at the almost unexplored 
range from 6 to 7 mm wavelength. The sensitivity attained over the whole 
frequency range (approximately 15\,mJy) allowed for the identification of 161 
spectral lines from 21 molecular species and 66 RRLs from hydrogen, helium, and 
carbon. The total number of isotopologues detected in the survey grows up to 
43, including the main species. Eight vibrationally excited species were also 
detected.

The spectrum of Orion KL at 7\,mm is dominated by maser emission from SiO, 
$^{29}$SiO, and $^{30}$SiO, which are mainly consistent with previous 
observations. In addition, we report the detection of a new intense feature at 
$-$1.4\kms\ in the $v$=2 SiO line. 

The large number of RRLs, up to $\Delta n=11$ is also remarkable. Velocities 
and line widths of the RRLs are consistent with those found at lower and higher 
frequencies, and the RRL emission is close to LTE. We have not found evidences 
for a previously reported change of RRL velocities with frequency, which has 
been claimed as the signpost of a density gradient within the {\sc Hii} region.

The molecular spectrum has been modelled by computing the emission of all the 
detected molecules but SiO (due to maser emission). The approach for the model 
is based on results previously gathered in similar surveys at 1, 2, and 3\,mm. 
This is the first radiative modelling of Orion~KL for eight of the detected 
molecules, three of them with only upper limits for the column densities.

The results and further modelling demonstrates that the Q-band (6 to 8\,mm 
wavelength) is especially sensitive to the low-temperature components in the 
environment of Orion~KL. Some complex organic molecules, such as CH$_3$CH$_2$CN 
and CH$_2$CHCN, seem to mostly arise from the hot core; CH$_3$NH$_2$, is 
probably in the same group, but is only observed at the detection limit in the 
present work. Moreover, although CH$_3$OCH$_3$ has been detected through lines 
at very low upper energy levels, the extended ridge component is not required 
to properly fit the observed line profiles. We also note the lack of a cold 
contribution to the emission of NH$_2$CHO and CH$_3$COCH$_3$, molecules well 
detected in the IRAM 30m survey. The model of warm components used to fit their 
lines at 3, 2, and 1.3\,mm properly constrains the emission of these species 
below the detection limit of the present work. Most of the remaining molecules 
seem to arise from both, the coldest and hottest parts of the source. A total 
of 18 spectral lines remain unidentified, two of them at a level above 5-sigma.

\begin{acknowledgements}
This work was done under the Host Country Radio Astronomy program at MDSCC. The 
authors wish to thank the MDSCC staff (especially the radio astronomy engineer) 
for their kind and professional support during the observations. This work is 
supported by MICINN (Spain) grants CSD2009-00038, and AYA2012-32032. J.R.R. 
acknowledges the support from project ESP2015-65597-C4-1-R (MINECO/FEDER). B.T. 
and J.C. also thank MINECO grants CTQ 2013-40717 P and CTQ 2010-19008, and the 
ERC for synergy grant ERC-2013-Syg-610256-NANOCOSMOS.
\end{acknowledgements}

\clearpage
\newpage
\clearpage

\clearpage
   \begin{figure*}
      \centering
      \includegraphics[width=0.96\textwidth]{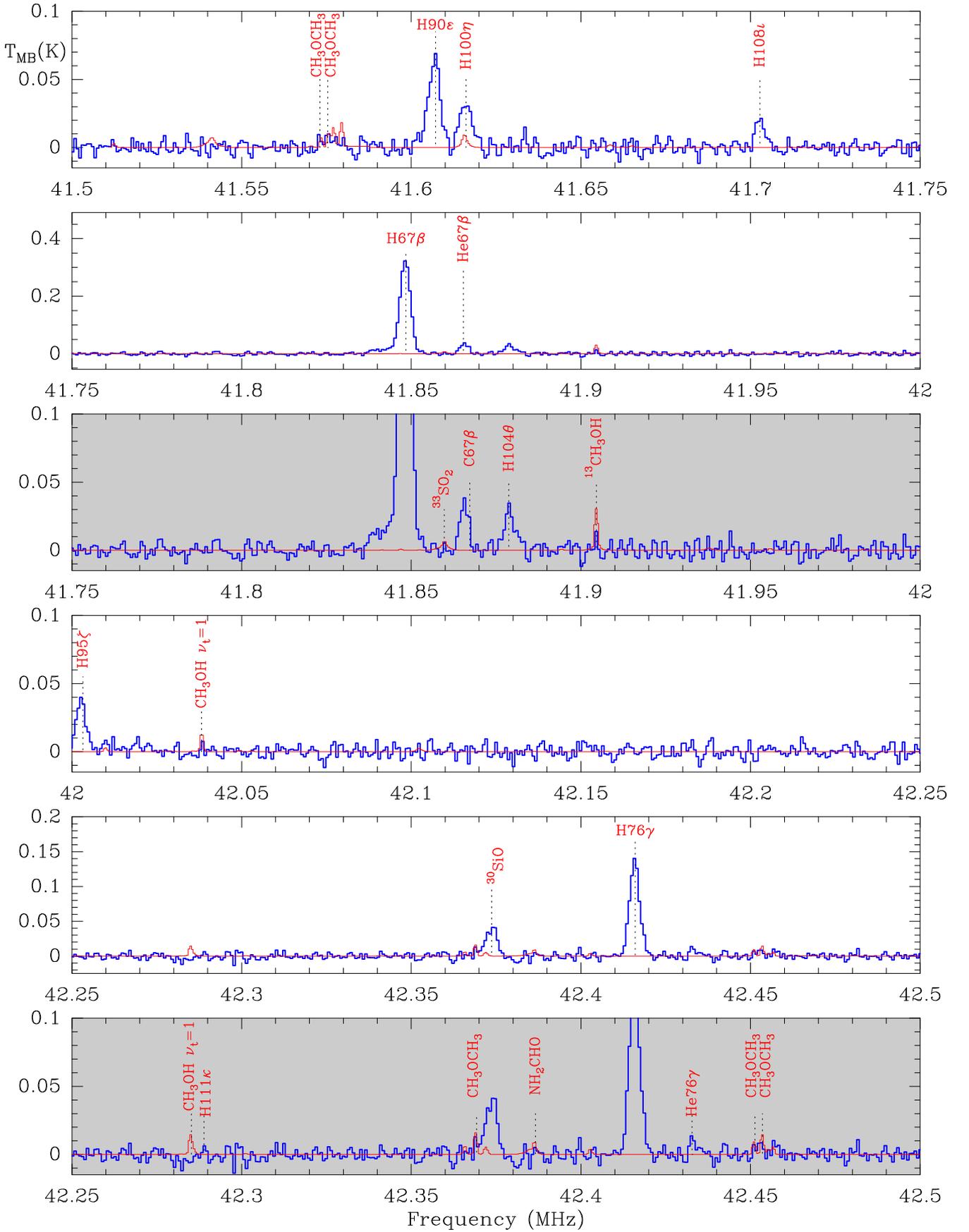}
      \caption{Detailed view of the observed spectrum of Orion KL at 7\,mm (in 
      blue), superimposed to the identification of the most intense lines 
      and the best fit model for the molecular species (in red). Each panel 
      displays a width of approximately 250\,MHz. In order to improve the 
      visibility of low intensity lines, some frequency ranges are plotted with 
      a zoom in intensity (grey boxes); a second level of zoom, when necessary, 
      is also depicted (yellow boxes). To compute the rest frequencies, a 
      velocity of $v_{\rm LSR}$ = +9\,km\,s$^{-1}$ is assumed. Note that the 
      model results do not include the SiO masers nor the RRLs.
              }
         \label{fig_model}
   \end{figure*}

   \begin{figure*}
      \includegraphics[width=0.96\textwidth]{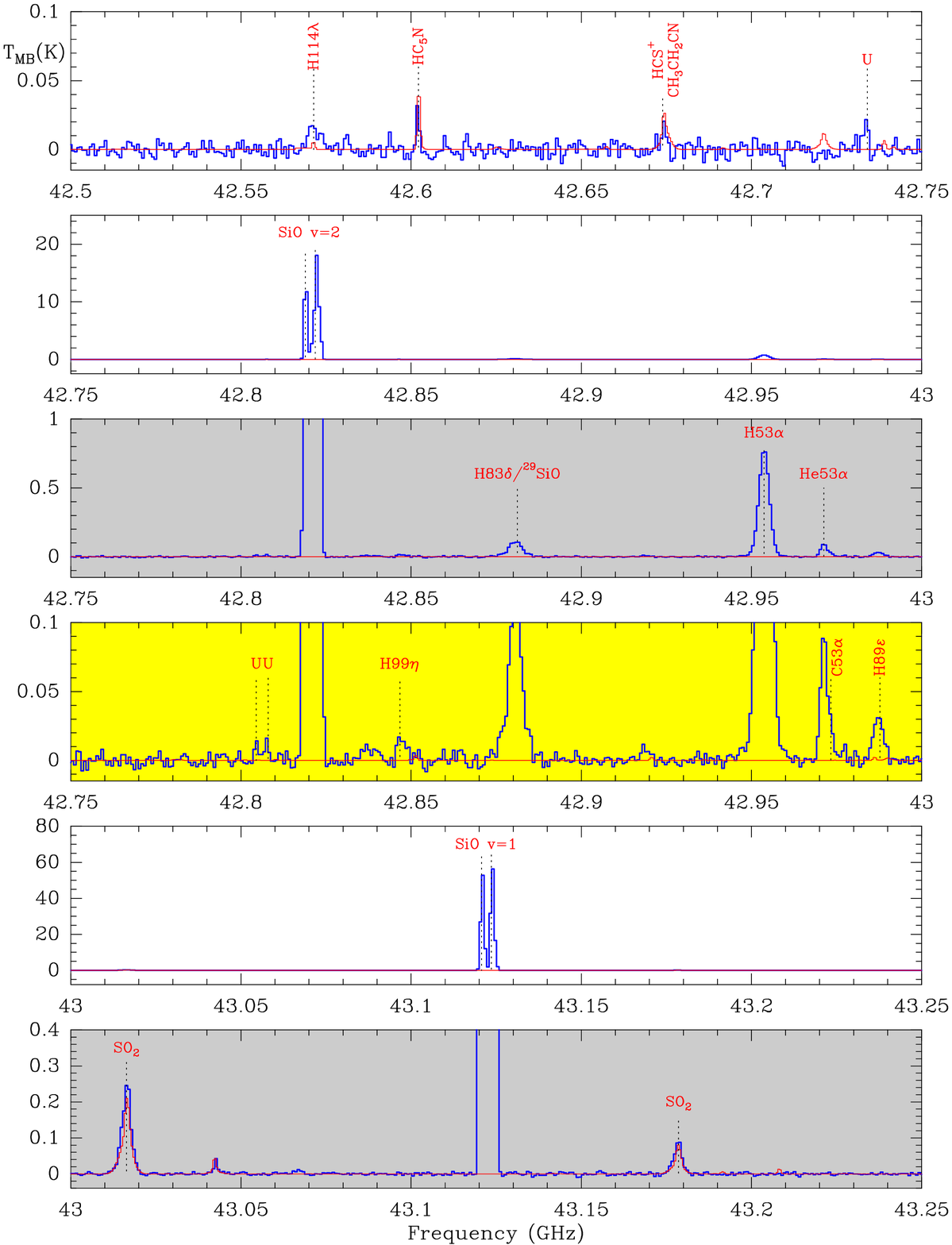}
      \\
      Figure \ref{fig_model}: continued.
   \end{figure*}

   \begin{figure*}
      \includegraphics[width=0.96\textwidth]{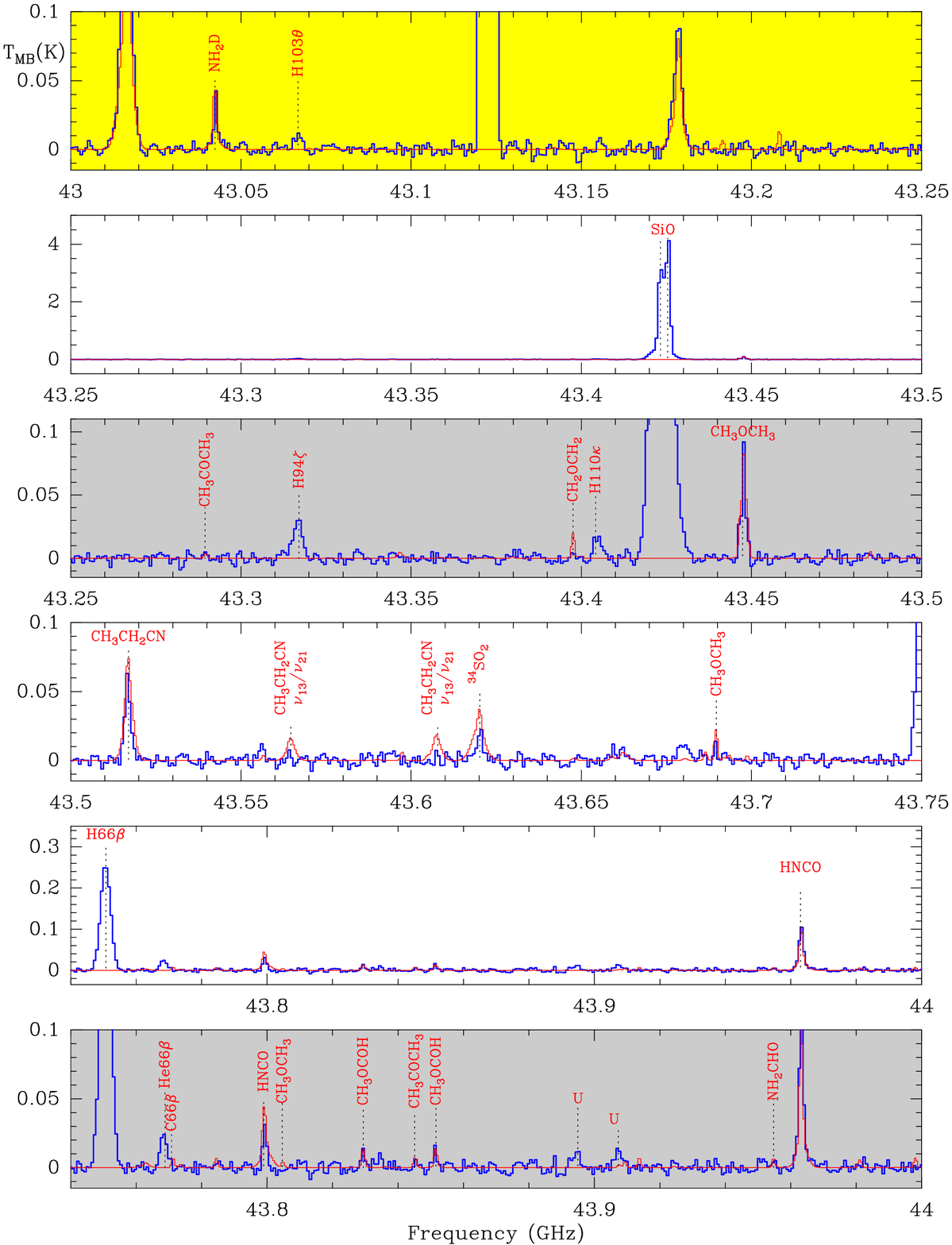}
      \\
      Figure \ref{fig_model}: continued.
   \end{figure*}

   \begin{figure*}
      \includegraphics[width=0.96\textwidth]{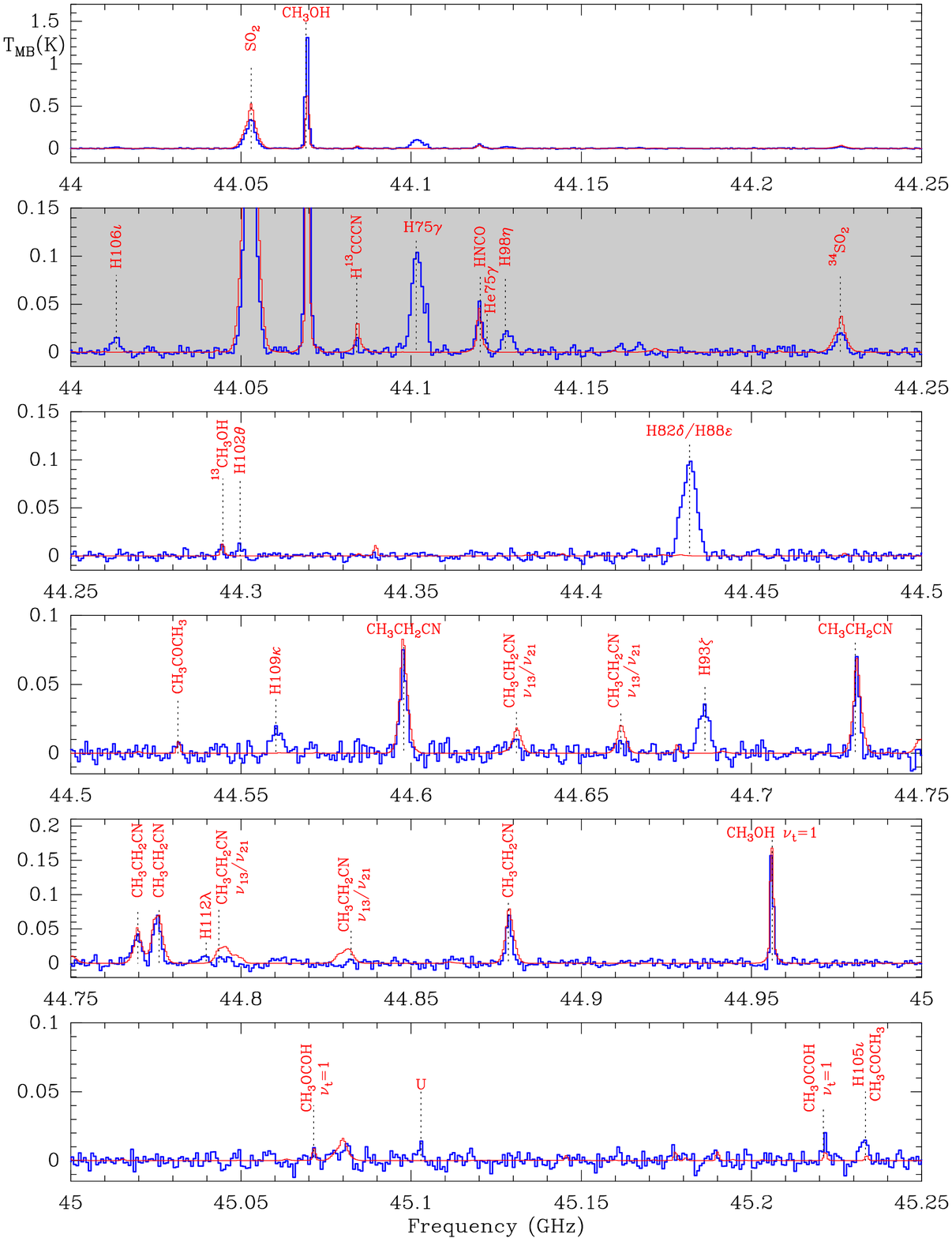}
      \\
      Figure \ref{fig_model}: continued.
   \end{figure*}

   \begin{figure*}
      \includegraphics[width=0.96\textwidth]{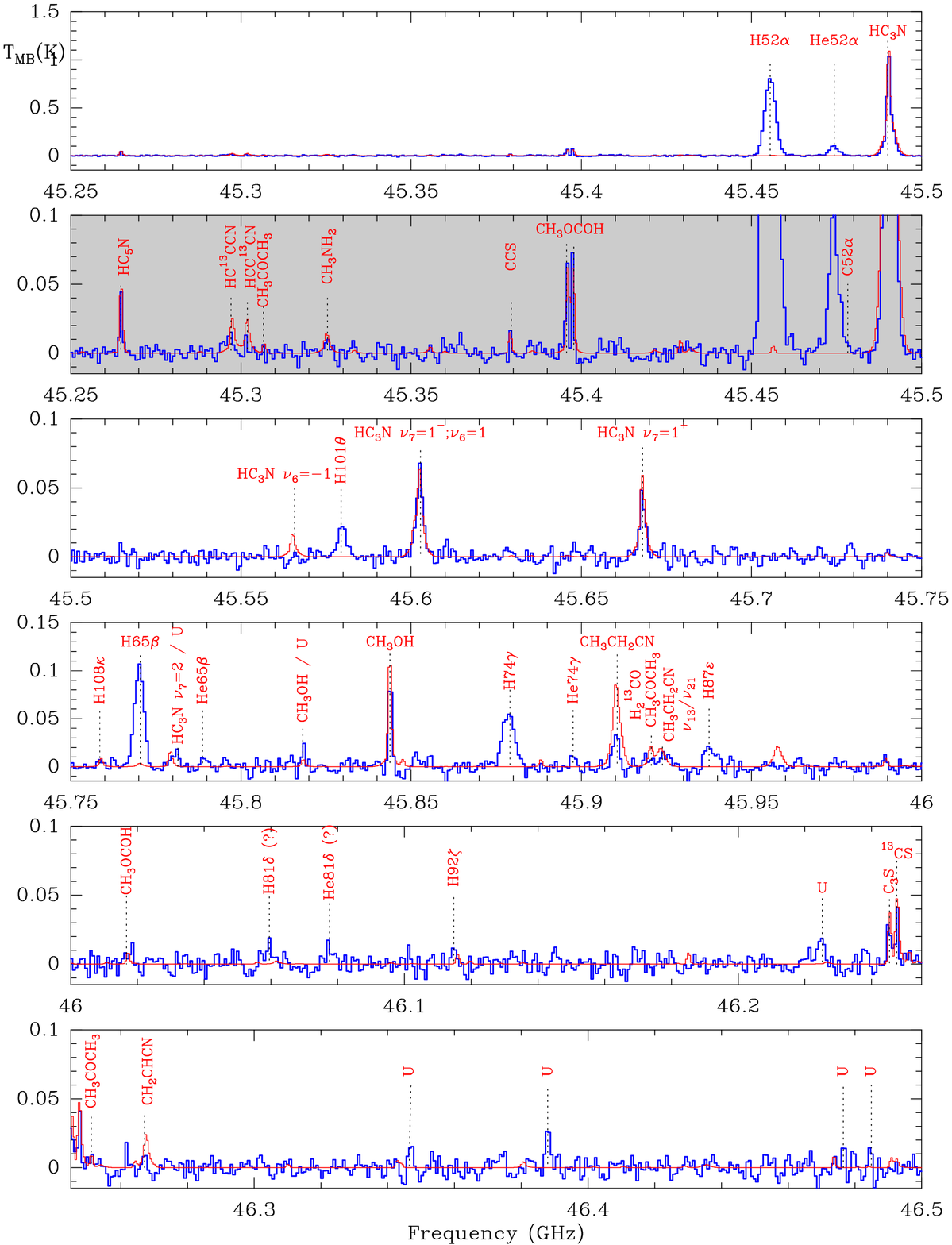}
      \\
      Figure \ref{fig_model}: continued.
   \end{figure*}

   \begin{figure*}
      \includegraphics[width=0.96\textwidth]{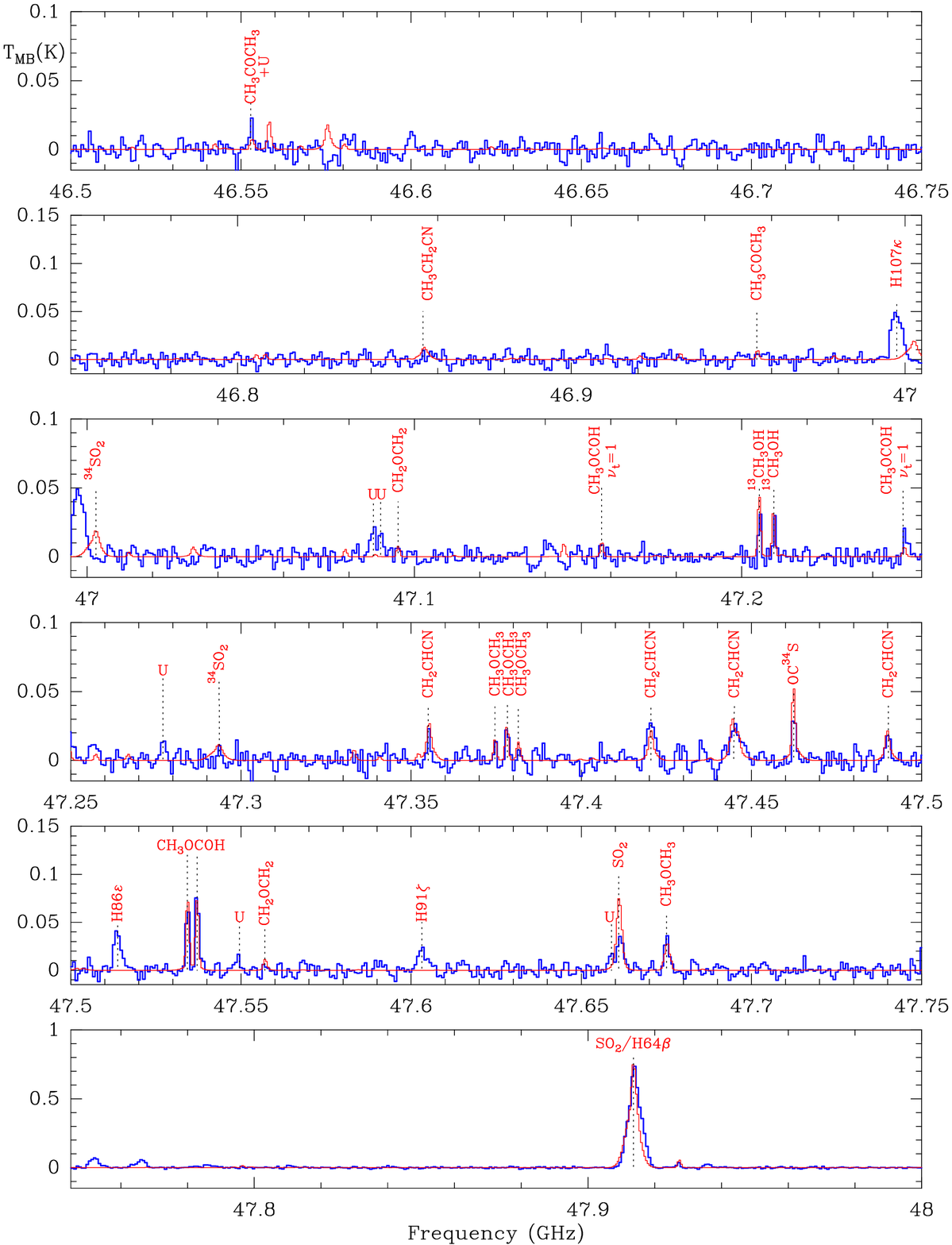}
      \\
      Figure \ref{fig_model}: continued.
   \end{figure*}

   \begin{figure*}
      \includegraphics[width=0.96\textwidth]{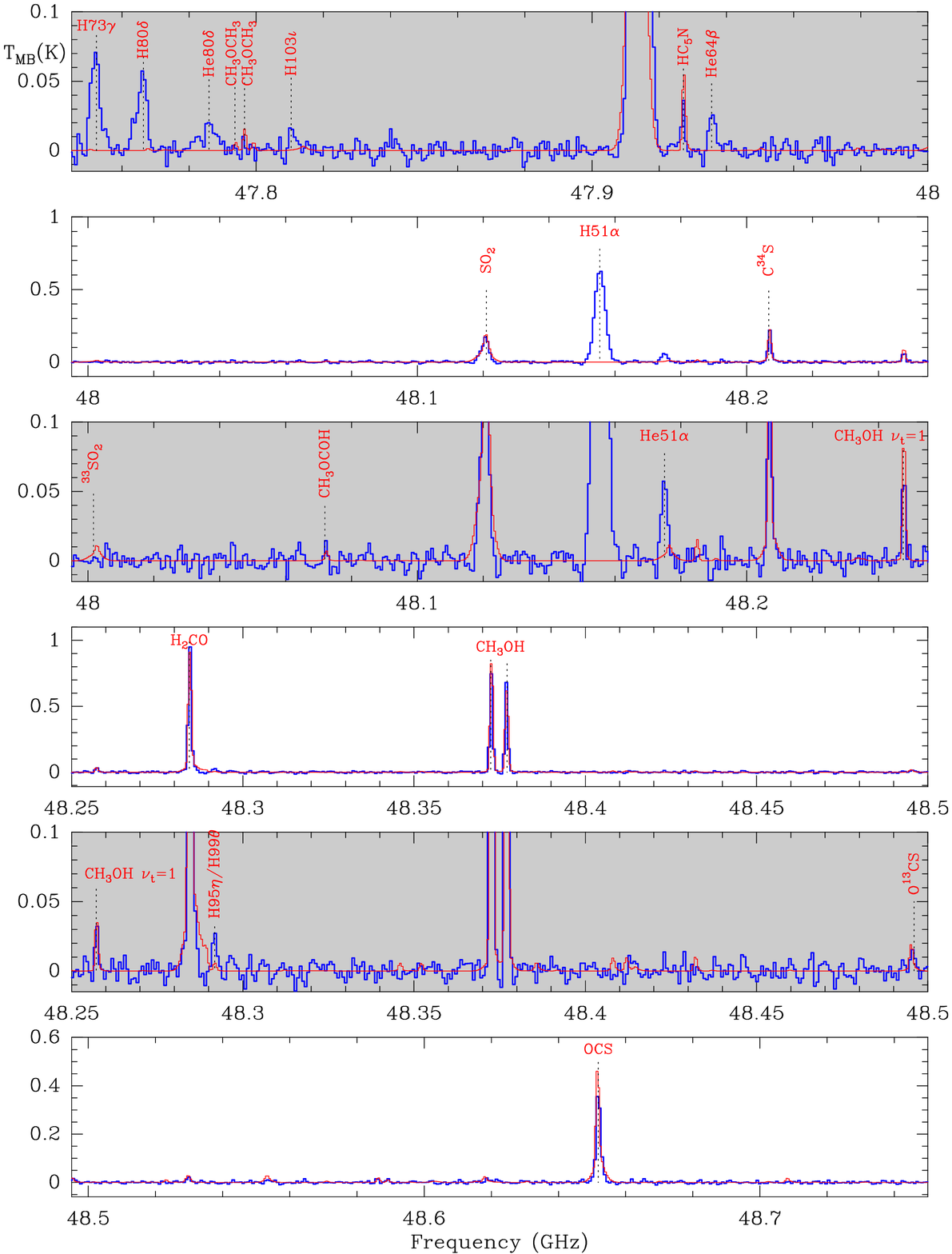}
      \\
      Figure \ref{fig_model}: continued.
   \end{figure*}

   \begin{figure*}
      \includegraphics[width=0.96\textwidth]{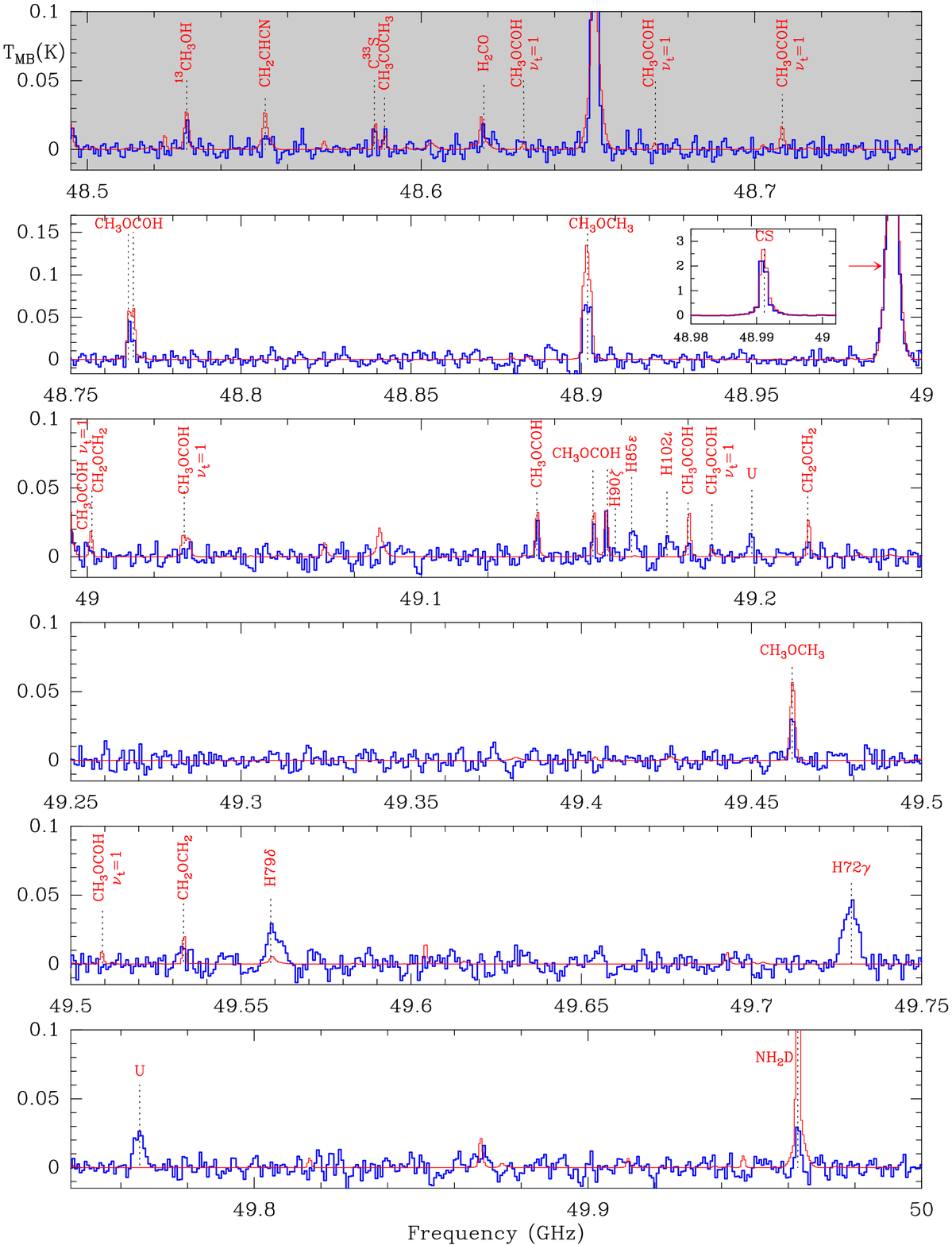}
      \\
      Figure \ref{fig_model}: continued.
   \end{figure*}

\clearpage
\newpage
\begin{table}
\begin{center}
\caption{DSS-54 antenna parameters \label{antenna}}
\begin{tabular}{lcr}
\hline\hline\noalign{\smallskip}
\multicolumn{2}{c}{Parameter} & \multicolumn{1}{c}{\qquad value} \\
\hline\noalign{\smallskip}

Diameter                &       $D$                     &       34~m \\
Beam size               &       $HPBW$                  &       \qquad 2050\arcsec/$\nu$(GHz) \\
Aperture efficiency     &       $\eta_{\mathrm A}$      &       0.46 \\
Main beam efficiency    &       $\eta_{\rm mb}$         &       0.52 \\
Sensitivity             &       $S/T_{\mathrm a}^*$     &       6.41 Jy/K \\

\noalign{\smallskip}\hline
\end{tabular}
\end{center}
\end{table}


\clearpage
\onecolumn

\longtab{

\begin{longtable}{lrrrrrl}        
\caption{Radio recombination lines detected \label{rrl}}\\
\hline\hline\noalign{\smallskip}                
\multicolumn{1}{c}{Line} & \multicolumn{1}{c}{Frequency} & \multicolumn{1}{c}{Area} & 
\multicolumn{1}{c}{Velocity} & \multicolumn{1}{c}{Width} & \multicolumn{1}{c}{Peak temperature} 
& \multicolumn{1}{c}{Notes} \\

& \multicolumn{1}{c}{MHz} & \multicolumn{1}{c}{K km s$^{-1}$} 
& \multicolumn{1}{c}{km s$^{-1}$} & \multicolumn{1}{c}{km s$^{-1}$} & \multicolumn{1}{c}{K} \\
\hline
\noalign{\smallskip}
\endfirsthead

\caption{continued.}\\
\hline\hline\noalign{\smallskip}                
\multicolumn{1}{c}{Line} & \multicolumn{1}{c}{Frequency} & \multicolumn{1}{c}{Area} & 
\multicolumn{1}{c}{Velocity} & \multicolumn{1}{c}{Width} & \multicolumn{1}{c}{Peak temperature} 
& \multicolumn{1}{c}{Notes} \\

& \multicolumn{1}{c}{MHz} & \multicolumn{1}{c}{K km s$^{-1}$} 
& \multicolumn{1}{c}{km s$^{-1}$} & \multicolumn{1}{c}{km s$^{-1}$} & \multicolumn{1}{c}{K} \\
\hline
\noalign{\smallskip}
\endhead
\hline

\\
\multicolumn{6}{c}{$\Delta\mathrm{n}=1$}  \\ \noalign{\smallskip}
H53$\alpha$	&	42951.9707	&	20.78  (0.07)	&	-3.29 (0.04)	&	24.82 (0.10)	&	0.787  (0.005)	&		\\
H52$\alpha$	&	45453.7217	&	20.97  (0.10)	&	-2.50 (0.05)	&	23.94 (0.13)	&	0.823  (0.007)	&		\\
H51$\alpha$	&	48153.6000	&	17.80  (0.10)	&	-2.76 (0.07)	&	24.24 (0.16)	&	0.690  (0.007)	&		\\
He53$\alpha$	&	42969.4730	&	 1.57  (0.11)	&	-3.24 (3.20)	&	15.30 (3.89)	&	0.096  (0.007)	&		\\
He52$\alpha$	&	45472.2435	&	 1.91  (0.34)	&	-2.65 (0.91)	&	16.92 (3.00)	&	0.106  (0.009)	&		\\
He51$\alpha$	&	48173.2220	&	 0.97  (0.08)	&	-2.41 (0.60)	&	14.16 (1.37)	&	0.065  (0.008)	&		\\
C53$\alpha$	&	42973.3984	&	 0.16  (0.11)	&	 8.74 (3.80)	&	5.34 (4.20)	&	0.028  (0.008)	&		\\
C52$\alpha$	&	45476.3975	&	 0.21  (0.33)	&	 8.46 (1.63)	&	6.76 (3.50)	&	0.030  (0.009)	&		\\
\\
\multicolumn{6}{c}{$\Delta\mathrm{n}=2$}  \\ \noalign{\smallskip}																						
H67$\beta$	&	41846.5533	&	 8.56  (0.10)	&	-2.92 (0.14)	&	24.85 (0.33)	&	0.323  (0.007)	&		\\
H66$\beta$	&	43748.9539	&	 6.47  (0.05)	&	-2.77 (0.09)	&	23.59 (0.21)	&	0.258  (0.004)	&		\\
H65$\beta$	&	45768.4478	&	 2.60  (0.07)	&	-2.17 (0.28)	&	21.57 (0.65)	&	0.113  (0.005)	&		\\
H64$\beta$	&	47914.1849	&	20.41  (0.14)	&	 9.96 (0.10)	&	28.47 (0.23)	&	0.674  (0.009)	&	(a)     \\
He67$\beta$	&	41863.6052	&	 0.82  (0.09)	&	-5.47 (1.01)	&	19.84 (2.22)	&	0.039  (0.006)	&		\\
He66$\beta$	&	43766.7811	&	 0.37  (0.04)	&	-2.26 (0.71)	&	14.33 (1.66)	&	0.024  (0.054)	&		\\
He65$\beta$	&	45787.0978	&	 0.14  (0.06)	&	-4.01 (3.76)	&	13.61 (7.21)	&	0.010  (0.006)	&		\\
He64$\beta$	&	47933.7093	&	 0.33  (0.08)	&	-3.97 (1.62)	&	12.63 (3.76)	&	0.025  (0.008)	&		\\
C67$\beta$	&	41867.4296	&	 0.07  (0.08)	&	10.71 (1.53)	&	5.41 (4.11)	&	0.013  (0.006)	&		\\
C66$\beta$	&	43770.7793	&	 0.02  (0.02)	&	 9.21 (0.95)	&	4.03 (2.07)	&	0.005  (0.004)	&		\\
\\
\multicolumn{6}{c}{$\Delta\mathrm{n}=3$}  \\ \noalign{\smallskip}																						
H76$\gamma$	&	42414.0908	&	3.57  (0.07)	&	-3.12 (0.22)	&	24.34 (0.54)	&	0.138  (0.005)	&		\\
H75$\gamma$	&	44100.0954	&	2.99  (0.07)	&	-3.13 (0.28)	&	27.09 (0.68)	&	0.103  (0.004)	&		\\
H74$\gamma$	&	45876.6701	&	1.55  (0.08)	&	-3.79 (0.69)	&	25.99 (1.50)	&	0.056  (0.006)	&		\\
H73$\gamma$	&	47749.9800	&	1.88  (0.17)	&	-3.69 (1.06)	&	24.30 (2.81)	&	0.073  (0.012)	&		\\
H72$\gamma$	&	49726.7007	&	1.10  (0.08)	&	-4.54 (1.02)	&	27.61 (2.16)	&	0.037  (0.006)	&	(b)	\\
He76$\gamma$	&	42431.3739	&	0.26  (0.14)	&	-2.43 (6.32)	&	15.56 (15.50)	&	0.016  (0.006)	&		\\
He75$\gamma$	&	44118.0656	&	0.59  (0.41)	&	-4.68 (5.27)	&	19.08 (17.10)	&	0.029  (0.025)	&	(c)     \\
He74$\gamma$	&	45895.3643	&	0.15  (0.05)	&	-5.10 (1.95)	&	11.52 (3.26)	&	0.012  (0.005)	&		\\
\\
\multicolumn{6}{c}{$\Delta\mathrm{n}=4$}  \\ \noalign{\smallskip}																						
H83$\delta$	&	42879.8047	&	3.87  (0.09)	&	 3.44 (0.35)	&	33.00 (0.89)	&	0.110  (0.005)	&	(d) 	\\
H82$\delta$	&	44430.7088	&	2.65  (0.04)	&	-0.24 (0.43)	&	26.19 (0.40)	&	0.095  (0.005)	&	(e)	\\
H81$\delta$	&	46057.3309	&	0.44  (0.11)	&	-4.33 (4.21)	&	30.43 (10.30)	&	0.014  (0.006)	&	(f)	\\
H80$\delta$	&	47764.3497	&	0.97  (0.08)	&	-2.93 (0.69)	&	17.20 (1.73)	&	0.053  (0.007)	&		\\
H79$\delta$	&	49556.7953	&	0.71  (0.11)	&	-9.13 (2.33)	&	28.05 (4.54)	&	0.024  (0.008)	&		\\
He81$\delta$	&	46076.0986	&	0.41  (0.08)	&	-1.23 (2.65)	&	27.29 (6.18)	&	0.014  (0.005)	&		\\
He80$\delta$	&	47783.8130	&	0.27  (0.10)	&	-6.19 (2.67)	&	16.62 (9.40)	&	0.015  (0.007)	&		\\
\\
\multicolumn{6}{c}{$\Delta\mathrm{n}=5$}  \\ \noalign{\smallskip}																						
H90$\epsilon$	&	41605.2445	&	2.14  (0.20)	&	-2.59 (1.31)	&	30.37 (3.36)	&	0.066  (0.012)	&		\\
H89$\epsilon$	&	42985.6946	&	0.79  (0.01)	&	-2.10 (1.13)	&	23.56 (2.55)	&	0.031  (0.005)	&	(g)	\\
H88$\epsilon$	&	44427.9131	&	1.03  (0.06)	&	-2.73 (0.92)	&	23.15 (0.95)	&	0.042  (0.005)	&	(h)	\\
H87$\epsilon$	&	45935.3946	&	0.86  (0.12)	&	-9.03 (2.34)	&	31.18 (5.53)	&	0.026  (0.008)	&		\\
H86$\epsilon$	&	47511.8739	&	0.56  (0.09)	&	-2.47 (1.06)	&	13.34 (2.28)	&	0.039  (0.009)	&		\\
H85$\epsilon$	&	49161.3453	&	0.30  (0.10)	&	-3.38 (2.33)	&	14.78 (5.48)	&	0.020  (0.009)	&		\\
\\
\multicolumn{6}{c}{$\Delta\mathrm{n}=6$}  \\ \noalign{\smallskip}																						
H95$\zeta$	&	42000.6430	&	0.80  (0.09)	&	-5.41 (1.08)	&	20.75 (2.56)	&	0.036  (0.006)	&		\\
H94$\zeta$	&	43314.7537	&	0.72  (0.06)	&	-4.50 (0.87)	&	22.12 (2.13)	&	0.031  (0.004)	&		\\
H93$\zeta$	&	44684.2804	&	0.95  (0.08)	&	-2.79 (1.03)	&	25.88 (2.28)	&	0.035  (0.005)	&		\\
H91$\zeta$	&	47601.5859	&	0.41  (0.09)	&	-1.60 (2.05)	&	19.07 (4.86)	&	0.020  (0.007)	&		\\
H90$\zeta$	&	49155.8585	&	0.12  (0.15)	&	-6.56 (8.02)	&	11.03 (14.50)	&	0.010  (0.017)	&	(i)	\\
\\
\multicolumn{6}{c}{$\Delta\mathrm{n}=7$}  \\ \noalign{\smallskip}																						
H100$\eta$	&	41613.9945	&	1.17  (0.10)	&	-6.00 (1.35)	&	32.52 (2.97)	&	0.034  (0.006)	&	(j)	\\
H99$\eta$	&	42845.8100	&	0.67  (0.07)	&	-4.64 (1.64)	&	28.80 (3.59)	&	0.022  (0.005)	&		\\
H98$\eta$	&	44126.7461	&	0.49  (0.06)	&	-2.23 (1.29)	&	22.07 (2.55)	&	0.021  (0.004)	&		\\
H97$\eta$	&	45459.2772	&	0.28  (0.08)	&	-3.32 (3.34)	&	22.46 (5.89)	&	0.012  (0.006)	&	(k)	\\
H95$\eta$	&	48289.7868	&	0.28  (0.58)	&	-3.57 (3.41)	&	8.35 (3.41)	&	0.031  (0.011)	&		\\
\\
\multicolumn{6}{c}{$\Delta\mathrm{n}=8$}  \\ \noalign{\smallskip}																						
H104$\theta$	&	41877.3884	&	0.60  (0.11)	&	-1.73 (1.78)	&	21.28 (5.16)	&	0.027  (0.007)	&		\\
H103$\theta$	&	43064.7499	&	0.20  (0.04)	&	-5.75 (1.83)	&	18.01 (3.40)	&	0.010  (0.003)	&		\\
H102$\theta$	&	44297.4483	&	0.26  (0.06)	&	-6.38 (1.62)	&	15.92 (4.84)	&	0.015  (0.004)	&		\\
H101$\theta$	&	45577.6694	&	0.38  (0.06)	&	-4.90 (1.37)	&	16.76 (3.43)	&	0.021  (0.005)	&		\\
H99$\theta$	&	48290.0704	&	0.26  (0.08)	&	-1.85 (1.31)	&	7.99 (2.95)	&	0.031  (0.010)	&		\\
\\
\multicolumn{6}{c}{$\Delta\mathrm{n}=9$}  \\ \noalign{\smallskip}																						
H108$\iota$	&	41700.7983	&	0.59  (0.07)	&	-4.73 (1.40)	&	23.37 (3.53)	&	0.024  (0.005)	&		\\
H106$\iota$	&	44011.4774	&	0.46  (0.06)	&	-3.96 (1.64)	&	25.13 (4.29)	&	0.017  (0.004)	&		\\
H105$\iota$	&	45231.0739	&	0.41  (0.07)	&	-4.25 (1.70)	&	19.22 (3.92)	&	0.020  (0.006)	&	(l) 	\\
H103$\iota$	&	47808.9329	&	0.16  (0.06)	&	-0.94 (2.37)	&	9.79 (3.71)	&	0.015  (0.007)	&		\\
H102$\iota$	&	49171.6054	&	0.29  (0.10)	&	-8.06 (3.82)	&	20.70 (8.41)	&	0.013  (0.009)	&		\\
\\
\multicolumn{6}{c}{$\Delta\mathrm{n}=10$}  \\ \noalign{\smallskip}																						
H111$\kappa$	&	42287.2908	&	0.11  (0.05)	&	-3.00 (3.02)	&	11.24 (5.23)	&	0.009  (0.006)	&		\\
H110$\kappa$	&	43402.8801	&	0.29  (0.05)	&	-4.60 (1.64)	&	16.91 (3.09)	&	0.016  (0.005)	&		\\
H109$\kappa$	&	44558.0836	&	0.38  (0.08)	&	-7.48 (2.67)	&	22.97 (5.14)	&	0.015  (0.006)	&		\\
H108$\kappa$	&	45754.6765	&	0.38  (0.09)	&	-4.61 (3.60)	&	29.54 (3.60)	&	0.012  (0.025)	&	(m)	\\
H107$\kappa$	&	46994.5305	&	1.12  (0.08)	&	-9.45 (0.79)	&	22.24 (1.66)	&	0.047  (0.006)	&	(n)	\\
\\
\multicolumn{6}{c}{$\Delta\mathrm{n}=11$} \\ \noalign{\smallskip}																						
H114$\lambda$	&	42569.5866	&	0.47  (0.08)	&	-4.87 (2.86)	&	29.11 (5.89)	&	0.015  (0.005)	&		\\
H112$\lambda$	&	44787.0910	&	0.20  (0.76)	&	-5.97 (3.38)	&	16.90 (6.71)	&	0.011  (0.007)	&		\\

\\
\hline
\\
\end{longtable}

\tablefoot{(a) Blend with SO$_2$ (47913.600 MHz). 2-components fitting impossible. 
(b) Probably contaminated by H98$\theta$.
(c) Blend with HNCO (44120.324 MHz). Separated. 
(d) Blend with $^{29}$SiO maser. 2-components fitting impossible.
(e) Blend with H88$\epsilon$. Separated. 
(f) Line clearly detected, but fitting not reliable due to spikes at close frequencies.
(g) Contaminated by CH$_3$COCH$_3$ (42985.958 MHz). 
(h) Blend with H82$\delta$. Separated.
(i) Close to CH$_3$OCOH doublet and H85$\epsilon$. 
(j) Probably contaminated by CH$_3$NH$_2$ (41615.387 MHz). 
(k) Very close to H52$\alpha$. Tentative fitting. 
(l) Probably contaminated by CH$_3$COCH$_3$ (45233.497 MHz). 
(m) Blend with CH$_3$CH$_2$CN (45758.528 MHz). Separated. 
(n) Blend with HNCS (46997.506 MHz). 2-components fitting impossible.}

}  


\clearpage
\onecolumn
\begin{table}
\begin{center}
\caption{LSR velocities of H$\alpha$ RRLs in Orion KL \label{Halpha-tab}}
\begin{tabular}{lrrrc}
\hline\hline\noalign{\smallskip}
\multicolumn{1}{c}{Line} & \multicolumn{1}{c}{\qquad Frequency} & 
\multicolumn{1}{c}{$V_{\rm LSR}$} & \multicolumn{1}{c}{Error} & 
\multicolumn{1}{c}{Ref.}\\
& \multicolumn{1}{c}{\qquad (GHz)} & 
\multicolumn{2}{c}{(km s$^{-1}$)} \\
\hline\noalign{\smallskip}

H30$\alpha$     &    231.9013     &     +4.0     &     1.3     &1\\
\\
H34$\alpha$     &    160.2115     &     -4.80    &     0.85    &2\\
H31$\alpha$     &    210.5018     &     -3.27    &     0.96    &2\\
H43$\alpha$     &     79.9127     &     -3.21    &     0.16    &2\\
H42$\alpha$     &     85.6884     &     -2.73    &     0.43    &2\\
H41$\alpha$     &     92.0344     &     -3.55    &     0.06    &2\\
H40$\alpha$     &     99.0230     &     -3.08    &     0.21    &2\\
\\
H71$\alpha$     &     17.9926     &     -5.4     &     0.1     &3\\
H70$\alpha$     &     18.7692     &     -4.4     &     0.1     &3\\
H69$\alpha$     &     19.5911     &     -3.4     &     0.1     &3\\
H68$\alpha$     &     20.4618     &     -3.6     &     0.1     &3\\
H67$\alpha$     &     21.3844     &     -5.1     &     0.1     &3\\
H66$\alpha$     &     22.3642     &     -4.9     &     0.1     &3\\
H64$\alpha$     &     24.5099     &     -3.6     &     0.1     &3\\
H63$\alpha$     &     25.6863     &     -2.6     &     0.1     &3\\
\\
H45$\alpha$     &     69.8296     &     -3.65    &     0.33    &4\\
H44$\alpha$     &     74.6446     &     -3.96    &     0.15    &4\\
H43$\alpha$     &     79.9127     &     -4.23    &     0.20    &4\\
H42$\alpha$     &     85.6884     &     -3.73    &     0.25    &4\\
H41$\alpha$     &     92.0344     &     -3.68    &     0.31    &4\\
\\
H53$\alpha$     &     42.9520     &     -3.29    &     0.04    &5\\
H52$\alpha$     &     45.4537     &     -2.50    &     0.05    &5\\
H51$\alpha$     &     48.1536     &     -2.76    &     0.07    &5\\

\noalign{\smallskip}\hline
\end{tabular}
\tablebib{
(1)~\citet{sut85}; (2)~\citet{ter10}; (3); ~\citet{gon15}
(4)~\citet{fra15}; (5)~this work.
}
\end{center}
\end{table}


\clearpage
\begin{longtab}
\begin{longtable}{lllcrrc}        
\caption{Line list \label{table_linelist}}\\
\hline\hline\noalign{\smallskip}                
\multicolumn{1}{c}{Observed} & \multicolumn{1}{c}{Species} & \multicolumn{1}{c}{Rest} & 
Transition & \multicolumn{1}{c}{$E_{\rm upp}$} & $A_{\rm ul}$ \\

\multicolumn{1}{c}{Freq. (MHz)} & \multicolumn{1}{c}{} & \multicolumn{1}{c}{Freq. (MHz)} 
& $J_{K_{\rm a},K_{\rm c}}$ & \multicolumn{1}{c}{(K)}  & $\times$10$^{-6}$ (s$^{-1}$)  \\
\hline
\noalign{\smallskip}
\endfirsthead

\caption{continued.}\\
\hline\hline\noalign{\smallskip}                
\multicolumn{1}{c}{Observed} & \multicolumn{1}{c}{Species} & \multicolumn{1}{c}{Rest} & 
Transition & \multicolumn{1}{c}{$E_{\rm upp}$} & $A_{\rm ul}$ \\

\multicolumn{1}{c}{Freq. (MHz)} & \multicolumn{1}{c}{} & \multicolumn{1}{c}{Freq. (MHz)} 
& $J_{K_{\rm a},K_{\rm c}}$ & \multicolumn{1}{c}{(K)} &  $\times$10$^{-6}$ (s$^{-1}$)  \\
\hline
\noalign{\smallskip}
\endhead
\color{red}{41573.1}          &   AA-CH$_3$OCH$_3$       &        41573.290 &  19$_{ 5,14}$-18$_{ 6,13}$          &  208.3  & 0.07  \\
\color{red}{41575.3}          &   EE-CH$_3$OCH$_3$       &        41575.465 &  19$_{ 5,14}$-18$_{ 6,13}$          &  208.3  & 0.06  \\
41607.1 &   H90$\epsilon$     &        41605.242 &              &        &   \\
41616.2 &   H100$\eta$           &        41613.994 &              &      &     \\
        &   CH$_3$NH$_2$           &        41615.387 &  15$_{ 1, 4}$-15$_{ 0, 5}$          &  256.9  & 0.86 \\
41702.7 &   H108$\iota$            &        41700.798 &           &     &      \\
41848.5 &   H67$\beta$     &        41846.551 &           &       &    \\
\color{red}{41859.7}  &   $^{33}$SO$_2$            &        41859.643 &   6$_{ 2, 4}$- 7$_{ 1, 7}$          &   29.0  & 0.14  \\
\hline
\end{longtable}
\tablefoot{$\dagger$: Observed frequency in red means that the line is below 3$\sigma$ (see Fig.\,\ref{fig_model}).}
\end{longtab}

\clearpage
\scriptsize
\begin{longtab}
\begin{longtable}{llccccclcc}        
\caption{Physical parameters derived by the models \label{table_model}}\\
\hline\hline\noalign{\smallskip}                
\multicolumn{1}{l}{Species} & \multicolumn{1}{l}{Component} & \multicolumn{1}{c}{$v_{\rm LSR}$} & 
\multicolumn{1}{c}{$\Delta$$v$} & \multicolumn{1}{c}{$d_{\rm sou}$} & \multicolumn{1}{c}{$T_{\rm K}$} & \multicolumn{1}{c}{$n$(H$_2$)}
 & \multicolumn{1}{c}{$N$} & \multicolumn{1}{c}{Method} & \multicolumn{1}{c}{Collisional}\\

\multicolumn{1}{c}{} & \multicolumn{1}{c}{} & \multicolumn{1}{c}{(km s$^{-1}$)} 
& \multicolumn{1}{c}{(km s$^{-1}$)} & \multicolumn{1}{c}{($''$)} &  \multicolumn{1}{c}{(K)} & \multicolumn{1}{c}{(cm$^{-3}$)} 
&\multicolumn{1}{c}{(cm$^{-2}$)} & \multicolumn{1}{c}{} & \multicolumn{1}{c}{rates}\\
\hline
\noalign{\smallskip}
\endfirsthead

\caption{continued.}\\
\hline\hline\noalign{\smallskip}                
\multicolumn{1}{l}{Species} & \multicolumn{1}{l}{Component} & \multicolumn{1}{c}{$v_{\rm LSR}$} & 
\multicolumn{1}{c}{$\Delta$$v$} & \multicolumn{1}{c}{$d_{\rm sou}$} & \multicolumn{1}{c}{$T_{\rm K}$} & \multicolumn{1}{c}{$n$(H$_2$)}
 & \multicolumn{1}{c}{$N$} & \multicolumn{1}{c}{Method} & \multicolumn{1}{c}{Collisional}\\

\multicolumn{1}{c}{} & \multicolumn{1}{c}{} & \multicolumn{1}{c}{(km s$^{-1}$)} 
& \multicolumn{1}{c}{(km s$^{-1}$)} & \multicolumn{1}{c}{($''$)} &  \multicolumn{1}{c}{(K)} & \multicolumn{1}{c}{(cm$^{-3}$)} 
&\multicolumn{1}{c}{(cm$^{-2}$)} & \multicolumn{1}{c}{} & \multicolumn{1}{c}{rates}\\
\hline
\noalign{\smallskip}
\endhead
CS              & Extended ridge & 9.0 & 4.0 & Extended   & 60  & 1$\times$10$^{5}$ & (5$\pm$2)$\times$10$^{14}$     & LVG & (a) \\
                & Plateau        & 6.0 & 25  & 30         & 125 & 1$\times$10$^{6}$ & (1.0$\pm$0.3)$\times$10$^{15}$ & LVG & (a) \\
                & Compact ridge  & 7.5 & 4.0 & 15         & 110 & 1$\times$10$^{6}$ & (5$\pm$2)$\times$10$^{15}$     & LVG & (a) \\
                & Hot core       & 5.5 & 10  & 10         & 225 &                   & (2.1$\pm$0.6)$\times$10$^{16}$ & LTE &     \\					
C$^{34}$S       & Extended ridge & 9.0 & 4.0 & Extended   & 60  & 1$\times$10$^{5}$ & (1.0$\pm$0.3)$\times$10$^{13}$ & LVG & (a) \\
                & Plateau        & 6.0 & 25  & 30         & 125 & 1$\times$10$^{6}$ & (8$\pm$2)$\times$10$^{13}$     & LVG & (a) \\
                & Compact ridge  & 7.5 & 4.0 & 15         & 110 & 1$\times$10$^{6}$ & (8$\pm$2)$\times$10$^{13}$     & LVG & (a) \\
                & Hot core       & 5.5 & 10  & 10         & 225 &                   & (7$\pm$2)$\times$10$^{14}$     & LTE &     \\
\hline
\end{longtable}
\tablefoot{(a): CS-He rates from \citet{Lique2007}; (b): Collisional rates derived from those of OCS \citep{Green1978} and from
the IOS approximation for a $^3$$\Sigma$ molecule (see \citealt{Corey1984,Corey1983}); (c): HCS$^+$-He rates from \citet{Monteiro1984};
(d): OCS-H$_2$ rates from \citet{Green1978}. (e): HC$_3$N-p-H$_2$ rates from \citet{wer07}. (f): HC$_5$N-H$_2$ rates estimate from
the desexcitation rates of \citet{deg84}.\\
($\dagger$): A+E species; ($\dagger\dagger$): AA+AE+EA+EE species.}
\end{longtab}
\twocolumn

\end{document}